\documentclass[12pt]{iopart}

\usepackage{graphicx}
\usepackage{epstopdf}
\usepackage{dcolumn}
\usepackage{bm}
\usepackage{rotating}
\usepackage[dvips]{color}
\begin{document}

\noindent{\it PREPRINT submitted to Reports on Progress in Physics}

\title{Formation and interactions of cold and ultracold molecules: new challenges for interdisciplinary physics}
\author{O Dulieu}
\address{Laboratoire Aim\'{e} Cotton, CNRS, B\^at. 505, Univ Paris-Sud 11,\\F-91405 Orsay Cedex, France}
\ead{olivier.dulieu@lac.u-psud.fr}

\author{C Gabbanini}
\address{Istituto per i processi chimico-fisici del C.N.R.,Via Moruzzi 1, 56124 Pisa, Italy}
\ead{carlo@ipcf.cnr.it}

\begin{abstract}
Progress on researches in the field of molecules at cold and ultracold temperatures is reported in this review. It covers extensively the experimental methods to produce, detect and characterize cold and ultracold molecules including association of ultracold atoms, deceleration by external fields and kinematic cooling. Confinement of molecules in different kinds of traps is also discussed. The basic theoretical issues  related to the knowledge of the  molecular structure, the atom-molecule and molecule-molecule mutual interactions, and to their possible manipulation and control with external fields, are reviewed. A short discussion on the broad area of applications completes the review.
\end{abstract}

\pacs{32.80.Pj, 33.80.Ps}
\maketitle



\tableofcontents

\section{Introduction}

"{\it Quo vadis, cold molecules?}". This was the title chosen by J. Doyle, B. Friedrich, F. Masnou-Seeuws, and R. Krems, for the editorial of the special issue on cold molecules published in 2004 in European Physical Journal D \cite{doyle2004}. This is still a very appropriate question nowadays, as the research field on cold  molecules (and atoms) is constantly expanding in many directions, involving a continuously increasing number of groups throughout the world. Cold molecules indeed open entirely new avenues for fascinating research, as it is brilliantly -sometimes "romantically"- expressed in several broad-scope papers. As early as 1998, J. Glanz reported on {\it the subtle flirtation of ultracold atoms} \cite{glanz1998}. "Molecules are cool" claimed J.~Doyle and B.~Friedrich \cite{doyle1999}, even "really cool" \cite{julienne2003}, also joined by B. Goss-Levi predicting "hot prospects for cold molecules" \cite{levi2000}. These "quantum encounters of the cold kind" lead researchers to see their "dreams of controlling interactions on the quantum level come true, and the exquisite nature of this control has proved remarkable" \cite{burnett2002}. This is actually the most prominent characteristic of this topic. For instance, cold molecules opened new perspectives in high-resolution molecular spectroscopy \cite{stwalley1999,jones2006,meerakker2005,gilijamse2007}. The expected accuracy of the envisioned measurements with ultracold molecules makes them appear as a promising class of quantum systems for precision measurements, related to fundamental issues usually discussed in the context of high-energy physics: the existence of the permanent electric dipole moment of the electron \cite{demille2000,hudson2002,kozlov2002,kawall2004} as a probe for CP-parity violation \cite{hinds1980,cho1991,ravaine2005}, the time-independence of the electron-to-nuclear and electron-to-nuclear mass ratios \cite{veldhoven2004,karr2005,schiller2005,chin2006,demille2008,zelevinsky2008}, and of the fine-structure constant \cite{hudson2006a}. Various proposals appeared for achieving quantum information \cite{demille2002,rabl2006,kotochigova2006,rabl2007}, or molecular optics devices \cite{kallush2005} based on cold polar molecules (i.e. exhibiting a permanent electric dipole moment). Elementary chemical reactions at very low temperatures could be manipulated by external electric or magnetic fields, therefore offering an additional flexibility for their control \cite{krems2005}. The large anisotropic interaction between cold polar molecules is also expected to give rise to quantum magnetism \cite{barnett2006}, and to novel quantum phases \cite{baranov2002,goral2002}. Last but not the least, the achievement of quantum degeneracy with cold molecular gases \cite{jaksch2002,donley2002,herbig2003,jochim2003,greiner2003,zwierlein2003,regal2003,bourdel2004} together with the mastering of optical lattices \cite{bloch2005} has built a fantastic bridge between condensed matter and dilute matter physics \cite{bloch2008}. Indeed, the smooth crossover between the Bose-Einstein condensation (BEC) of fermionic atomic pairs and the Bardeen-Cooper-Schrieffer (BCS) delocalized pairing of fermions related to superconductivity and superfluidity, has been experimentally observed \cite{bourdel2004,chin2004,bartenstein2004,zwierlein2004}. It is also expected that polar molecules trapped inside an optical lattice could provide lattice-spin models \cite{micheli2006}.

Due to the specific domain covered by the present review, it is first worthwhile to precisely define the basic words of its topic, i.e. {\it molecule} and {\it cold}, as most readers are probably not experts in this field.

\subsection{What do we mean by a "molecule"?}

As it will become clear further on, we adopt a rather broad definition for a molecule, being a {\it tightly or loosely} bound collection of a small number of atoms, held together by the electromagnetic field of the constituent atoms, or even by applied external electromagnetic fields. We will mostly focus on diatomic or triatomic molecules, except otherwise stated \footnote{Indeed, most of the experimental realizations of cold molecular samples originate from association of cold atom pairs, or from slowing down of simple molecules}. This definition obviously includes the standard vision of the strong chemical bond inside molecules in a stable state (the "chemist's molecules"), induced by electron exchange interaction responsible for many collision processes or elementary chemical reactions. The molecular ground state binding energy is typically in the 0.1-10~eV range, while the molecular equilibrium distance most often lies between 2 and 10$a_0$  \footnote{In this paper, we will mostly use atomic units (a.u.) for distances, with 1~a.u. $\equiv a_0$=0.0529177~nm, and wave numbers (cm$^{-1}$) as energy units, defined by two times the Rydberg constant $2R_{\infty}=217474.63137$~cm$^{-1}=27.2116$~eV}. Another class of well-studied {\it stable} molecules is composed by the Van der Waals molecules \cite{demtroeder1998}, most often involving rare gas atoms, which exist thanks to the mutual polarization of their individual atomic or molecular components. This is a long-range interaction whose potential energy varies as $R^{-6}$ ($R$ being the distance between individual components), leading to binding energies in the 0.01-0.1~eV range or below, and equilibrium distances in the 10$a_0$ range. Another kind of molecular bond arises from the competition between the ion-pair configuration (A$^+$+B$^-$) and the covalent configuration (A+B) within electronically {\it excited} states of various diatomic molecules AB. This feature generally produces a double-well pattern on the relevant potential curves, the inner one being related to the conventional chemical bond, while the position and depth of the outer depends on the strength of this competition. Among well-known examples are for instance the $B"\bar{B}$ state in H$_2$ \cite{sidis1983,wolniewicz1988}, or highly-excited states of alkali dimers \cite{laue2003a}.

\begin{figure}
\begin{center}
\includegraphics[width=0.8\textwidth]{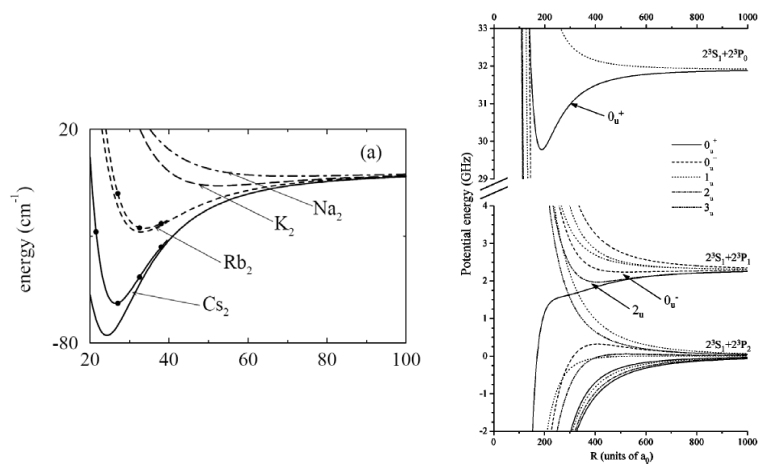}
\end{center}
\caption{Long-range potential wells in the excited molecular states of several diatomic molecules. (a) The so-called $0g^-$ states in alkali dimers, correlated to the lowest $^2S_{1/2}+^2P_{3/2}$ dissociation limit (reprinted with permission from ref.\cite{fioretti1999}). (b) The so-called $0_u^+$ state in metastable helium dimer, correlated to the $2^3S+2^3P$ dissociation limit (reprinted with permission from ref.\cite{leonard2004}). In both cases, note the range of internuclear distances and of potential energies, respectively much larger and much smaller than those of the "chemist's molecules".}
\label{fig:longrangewells}
\end{figure}

An important achievement of the cold molecule researches in the 90's has been the experimental observation of a new kind of {\it excited} electronic molecular states, usually referred to as {\it pure long-range molecules} in the literature \cite{jones1996,molenaar1996,amelink2000,wang1997,miller1993,cline1994,amiot1995,fioretti1999,jelassi2006a,fioretti2001} which have been predicted thirty years ago \cite{movre1977,stwalley1978} for most homonuclear alkali dimers (from Na$_2$ to Cs$_2$). In a few cases, the competition between the long-range dipole-dipole interaction of a ground state $S$ atom with an excited $P$ atom of the same species (which varies as $R^{-3}$) with the spin-orbit interaction creates a double-well structure in the related molecular excited potential curves: the standard inner well for chemical bonding is now accompanied by a tiny outer well located at large distances. Inside the bound pair, the atoms are left apart at distances of several tens of Bohr radii (Figure \ref{fig:longrangewells}a), with binding energies ranging from a fraction of cm$^{-1}$ for Na$_2$ up to a few tens of cm$^{-1}$ in Cs$_2$. The vibrational levels of such excited states generally decay within a few tens of nanoseconds into a pair of free ground state atoms, but they live long enough to be well characterized spectroscopically. The helium dimer formed by a metastable $2^3S$ atom and an excited $2^3P$ atom is an even more spectacular case \cite{leonard2004,leonard2005}, exhibiting an equilibrium distance as large as 200~$a_0$, with a binding energy as low as 0.05~cm$^{-1}$ (Figure \ref{fig:longrangewells}b). Jones {\it et al} \cite{jones2006} refer to this class of molecules as {\it physicists' molecules}, i.e., molecules whose properties can be related with high precision to the properties of the constituent atoms. In particular, such molecular states have been proven useful for the accurate determination of atomic radiative lifetimes (see for instance the review of ref.\cite{bouloufa2008}). As described in the next section, these advances benefited from the extensive studies of the {\it photoassociation} process between laser-cooled atoms \cite{stwalley1999,jones2006,weiner1999,masnou-seeuws2001,dulieu2003}, i.e. where two colliding cold atoms resonantly absorb a photon and produce an excited molecule in a well-defined rovibrational level.

It is recognized for a while now that the hyperfine interaction, i.e. the coupling between electronic and nuclear angular momenta, plays a crucial role in the photoassociation of cold atoms, especially when looking at molecular energy levels close to the dissociation limit related to a ground state $S$ atom and an excited $P$ atom of the same species \cite{kerman2004,tiesinga2005}. Pure-long range molecular states are also predicted as resulting from the competition between the dipole-dipole and the hyperfine interaction, just like in the case above. However, they have not been observed up to now, due to the interplay of the hyperfine interaction in the excited state of alkali atoms with their broad natural width. In contrast, such states can be probed in the $^{171}$Yb dimer through the observation of photoassociation transition close to the $^1S_0-^3P_1$ intercombination line with a very small natural width \cite{enomoto2008}. The shallowest long-range potential well ever identified has been observed, as deep as 750~MHz, about half of the depth of the long-range potential well of the helium dimer \cite{leonard2004}.

During the same period the electronic ground state of alkali diatomic molecules has been explored up to its dissociation limit, as stable molecules have been created in the uppermost ro-vibrational levels bound by a tiny energy much smaller than a cm$^{-1}$. Again, the molecular bond is peculiar: most of the vibrational motion takes place at unusually large distances (typically about 100$a_0$), in a region where the Van der Waals interaction varying as $R^{-6}$ binds the atoms together. But in contrast with the Van der Waals molecules, or with the pure long-range molecules above, the atoms can be accelerated towards each other inside the chemical bond region where they stay for a short time (less than 1~ps), while the total vibrational period can reach values as high as 1~ns. In other words, the potential energy of the atom pair may vary by about eight orders of magnitude during a single vibrational period. These molecules represent an even more striking example of physicists' molecules. Such high-lying molecular levels result for instance from the spontaneous decay of those pure long-range molecules excited by photoassociation. At such distances, the electrostatic interaction competes with the hyperfine interaction giving rise to resonant patterns in the dynamics of the atom pair, known as Feshbach resonances \cite{kohler2006}. The control of such resonances with external magnetic fields yielded a novel way to associate cold atom pairs into molecules (usually referred to as magnetoassociation or Feshbach association), and opened the amazingly creative research field of quantum degeneracy with molecules. In particular, the universality of the few-body physics has been discussed in the {\it quantum halo} regime \cite{jensen2004}, where the size of a bound atom pair exceeds the range of the atom-atom interaction, so that its wave function extends far inside the classically-forbidden region. A well-known example of such a system is the ground state of the helium dimer $^4$He$_2$ \cite{tang1995}. Efimov states, i.e. bound states of three particles are predicted in this regime \cite{efimov1990,efimov1993}.

To conclude this paragraph, it is worth mentioning an even more exotic class of {\it excited} molecules, identified as long-range Rydberg molecules \cite{greene2000,hamilton2002,khuskivadze2002}. In one case, a highly-excited atom A$^*(n \ell)$ (a Rydberg atom) can be associated with a neighboring ground state atom A to give rise to a tiny bound molecular state, due to the interaction between the quasi-free electron of A$^*$ (with principal quantum number $n$ and angular momentum $\ell$) and the neutral atom which acts as a perturber. The size of the resulting molecule is comparable to the size of the Rydberg atom, which scales as $n^3$, and can reach thousands of $a_0$, i.e. the size of a bacteria. Moreover, such molecules, while homonuclear, should exhibit a giant permanent electric dipole moment (with a magnitude in the range of $10^3$~Debye) just like Rydberg atoms, and lifetimes of the order of hundreds of microseconds. In a second case, pairs of Rydberg atoms are predicted to form "macrodimers", or ultralong-range Rydberg molecules, bound by the long-range electrostatic interaction between them \cite{boisseau2002}. Such Rydberg molecules of either type have a typical size in the $10^3-10^4 a_0$ range, and a binding energy far smaller that a cm$^{-1}$. Experimental evidence for the existence of such peculiar systems has been recently reported in the literature \cite{farooqi2003,singer2004,greene2006,bendkowsky2009}.

In the rest of this paper, we will focus on molecules in their ground electronic state, or in their lowest metastable state, when appropriate.

\subsection{What do we mean by "cold" and "ultracold"?}

Many of the systems or phenomena quoted above have been investigated in the context of what is nowadays usual to refer to as {\it cold} or {\it ultracold} matter. In the common sense, cold is just "cold as ice". In the thermodynamical sense, the quest of low temperatures matches the researches on gas liquefaction, down to the ultimate limit of the liquid $^3He$ at about 2 millikelvins (mK). Since the late eighties, the amazing development of laser cooling of dilute atomic gaseous samples \cite{phillips1998,chu1998,cohen-tannoudji1998} established the current nomenclature for the field of cold and ultracold matter, respectively associated to a temperature above, and below 1~mK. Nevertheless, in many physical situations in this research area, the system under study is not in thermodynamical equilibrium, and its temperature $T$ should be understood as resulting from the identity $E_{kin} \approx k_BT$, where $E_{kin}$ is the representative kinetic energy of the constituting atoms and/or molecules ($k_B$ being the Boltzmann constant). As typical examples, a dilute gas of laser-cooled cesium atoms in a magneto-optical trap (MOT), with density $n \approx 10^{11}$~atoms/cm$^3$, is created almost at thermodynamical equilibrium with $T \approx 0.1$~mK, while the residual motion of cesium atoms in a Bose-Einstein condensate (BEC) corresponds to a temperature of the order of a few nanokelvins (nK). But the supplementary degrees of freedom of molecules, i.e. vibration and rotation, compared to atoms, result into their most fascinating properties as yielding further possibilities to manipulate them, and this is the central topic of this review paper. Therefore it is suitable to define a temperature for each of these degrees of freedom, say $T_t$, $T_v$ and $T_r$ for the translational, vibrational, and rotational degrees of freedom, respectively, just like for instance in molecular thermal or supersonic beams. In the various physical situations quoted above, it is clear that translationally cold or ultracold conditions are required to be able to observe those tiny bound molecular systems, otherwise they would be destroyed immediately by collision with surrounding particles. In most cases, translationally cold molecules created by photoassociation or magnetoassociation do not rotate, i.e. are rotationally cold. In contrast they are most often created in  a broad distribution of vibrational levels (vibrationally hot) so that they contain an enormous amount of internal energy which can be released even during a cold collision, which blows out the entire system. The control of this energy release is currently the main concern of many researches in the cold molecule community, as we will see below.

\subsection{Is there a universal approach to create cold and ultracold stable molecules?}

This will be the central question of the next section, and we will see that every class of approaches has its own degree of "universality". At first it is natural to imagine that laser-cooling could be used for molecules just like for atoms to bring them almost at rest. But the complex internal structure of the molecules generally prevents this approach to work \cite{bahns1996}. Indeed, laser-cooling of atoms relies on a closed-level radiative transition scheme which allows the atoms to undergo many absorption-spontaneous emission cycles until they are stopped in the laboratory frame. Such closed-level scheme are not expected to be easily found in molecules \cite{dirosa2004,stuhl2008}, so that the population is spread over many rovibrational levels after a few absorptions. Very soon came the idea to associate a pair of ultracold atoms using electromagnetic fields (photoassociation) \cite{thorsheim1987}, which does not change the translational motion of the pair, then creating a stable ultracold molecule. The formation of ultracold molecules using photoassociation in a MOT of ultracold atoms has been observed for the first time with cesium atoms \cite{fioretti1998}. The observation of even colder diatomic molecules has been reported after a two-photon Raman transition inside a rubidium BEC \cite{wynar2000}. The possibility to use external magnetic field to associate ultracold atoms (magnetoassociation) has been first pointed out in refs.\cite{timmermans1999,heinzen2000}, based on the important concept of magnetic tunability of Feshbach resonances to control interatomic interactions \cite{tiesinga1993}. Their experimental observation can be considered as having been initiated by the first demonstration of an atom-molecule coupling in a rubidium BEC \cite{donley2002}. But it is important to mention that these methods are by far not universal, as being restricted mostly to the formation of {\it ultracold} ground state {\it alkali dimers} in high-lying vibrational levels.

A somewhat more intuitive idea could be to use cryogenics methods to reach low temperatures. Indeed it is possible to set up a cryogenic cell with liquid helium at temperature well below 1~K, producing a helium vapor as well. The group of J. Doyle has promoted the technique of buffer-gas cooling, first on atoms like europium \cite{kim1997} and chromium \cite{weinstein1998}, and next on the CaH molecule \cite{weinstein1998a}. As long as a molecular gas can be introduced in the cell -for instance by laser ablation on a solid target- the molecules are thermalized by collisions with the surrounding cold helium gas. They can be trapped in a magnetic trap while the buffer gas is pumped out from the cell. In principle this technique can be applied to any paramagnetic molecule. An alternate experimental approach actually achieved a microscopic "flying version" of the buffer-gas cooling technique: molecular species can be isolated in the isothermal environment ($T \sim 0.15-0.37$~K) provided by helium droplets (containing thousands to billions of atoms) from an helium supersonic beam \cite{toennies2004}. The weak interaction between the trapped species and the helium droplet has allowed many spectroscopic investigations with molecular species, as the embedded molecule can rotate freely \cite{goyal1992,stienkemeier1995}.

Another option is to slow down existing molecules using external fields. The Stark deceleration of molecular beams via inhomogeneous electric field is a powerful technique pioneered in the group of G. Meijer \cite{bethlem2003,yamakita2004,vanhaecke2005}, and certainly represents a breakthrough in the manipulation of polar molecules, until their storage in various kinds of traps \cite{bethlem2000,hoekstra2007,heiner2007}. One can also exploit electrostatic field \cite{rangwala2003,rieger2005} or magnetic field \cite{patterson2007} to select and guide slow molecules out of a room-temperature reservoir . Intense optical fields can also be used to create traveling optical lattices for deceleration purpose \cite{fulton2004,fulton2006}. A new emerging idea consists in implementing analogous approaches using pulsed magnetic fields to slow down paramagnetic molecules \cite{vanhaecke2007,hogan2007,hogan2008}, with a first success already demonstrated with molecular oxygen \cite{narevicius2008}. All these approaches have a limited applicability: they are restricted to {\it light neutral molecules or radicals}, which are slowed down to kinetic energies equivalent to temperatures above the millikelvin, i.e. the {\it cold} domain.

A couple of other approaches have also been demonstrated, relying on {\it kinematic} cooling of molecular systems, which are sometimes considered as the "most universal" ones as they rely on purely kinematic considerations, while not so easy to handle. Hence the collision of an atomic (say, Ar) and a molecular (say, NO) beam in an appropriate geometric configuration can leave one of the species (NO) at rest in the lab frame, the excess of energy and momentum being carried out by the other species (Ar) \cite{elioff2003}. A backward rotating nozzle filled with oxygen has been shown to deliver pulses of O$_2$ molecules at velocities equivalent to 10~K \cite{gupta2001}. We also note that the possibility to cool rotational, vibrational, and translational degrees of freedom of molecules by coupling a molecular dipole transition to an optical cavity has been predicted \cite{vuletic2000} and simulated \cite{morigi2007,kowalewski2007}.

Finally, we will not cover the topic of cold molecular ions in this review. This is a very active area nowadays, since the first observation of the formation of molecular ions inside a trap of laser-cooled magnesium ions \cite{molhave2000}. This area is nicely covered in a recent book edited by I. W. M. Smith \cite{smith2008}.

\section{The challenge for experimentalists: obtaining cold molecules}

As emphasized above, several routes have been demonstrated to create samples of cold molecules under various conditions of density and temperature, often related to their chemical nature. Indeed, laser techniques will be adapted to species for which resonant laser frequencies are available, i.e. are mostly applied to cold alkali, alkaline-earth, and metastable rare gas atomic samples. In contrast, deceleration of preexisting molecules by an external electric or magnetic field will be used for lighter species due to experimental constraints on the length of their flying area. We review in this section the main investigations performed during the last fifteen years, with the goal of creating dense ensembles of cold {\it neutral} molecules in their absolute ground state.


\subsection{Association of ultracold atoms: the quest for ultracold molecules in their absolute ground state}

In the last few years researchers have learned how to use the huge advances in laser cooling, trapping, and manipulation of atomic gases to assemble molecules from ultracold atoms. Two main processes can be used to this aim: (i) photoassociation, where two colliding cold atoms can be fused together through the absorption of a photon into an excited molecule which is stabilized by spontaneous emission; (ii) magnetoassociation, based upon the existence of Feshbach resonances arising from the coupling between the initial scattering state of the free atoms and of a bound molecular state. In the following subsections we describe both methods and review the related experiments. In most cases, molecules are created in high vibrational states which are unstable against vibrational relaxation through collisional or radiative processes. In the last subsection we present the most recent experimental developments devoted to the method for obtaining samples of cold molecules in their absolute ground state.

\subsubsection{Molecules formed after photoassociation of cold alkali atom pairs}
\label{sssec:pa}

Photoassociation (PA) occurs when two colliding laser-cooled atoms resonantly absorb a photon and produce an excited molecule in a given rovibrational level ~\cite{thorsheim1987}. During the last two decades, PA of cold atoms appeared as a powerful tool for high-resolution molecular spectroscopy \cite{stwalley1999}, giving access to the detailed knowledge of the long-range interactions between atoms in their ground or excited state. For instance, information about atomic collisional parameters at very low energies, such as scattering lengths and Feshbach resonance positions, has been determined. As the kinetic energy of the cold colliding atoms is very well defined ($E_{\mathrm{kin}}/ k_{\mathrm{B}} \leq 1$~mK, corresponding to about 20~MHz), PA is a quasi-resonant process, and high-resolution spectra of excited molecular states can be recorded as a function of the PA laser frequency $\omega_{\mathrm{PA}}$. The efficiency of the PA process depends on the density of atomic pairs at internuclear distance $R$, which scales as $R^{2}$, so it probes preferentially excited vibrational levels close to the corresponding dissociation limit, i.e. with a large spatial extension. The PA rate strongly depends on the detuning $\Delta = \omega_{\mathrm{PA}}-\omega_{\mathrm{at}}$ from the atomic resonance frequency $\omega_{\mathrm{at}}$. For a pair of identical alkali atoms where the asymptotic part of the excited state potential is determined by the resonant dipole-dipole interaction $V(R) = D-C_{3}/R^{3}$, it scales as $\Delta^{-7/6}$ \cite{pillet1997,cote1998}. However PA spectroscopy turned out to be efficient even down to several hundreds of cm$^{-1}$ below resonance, and thus appeared as complementary to the conventional molecular spectroscopy in determining the complete potential curves of molecules. Precise determinations of the $C_{3}$ coefficient yielded accurate values of the radiative lifetime of the lowest excited level of alkali atoms \cite{mcalexander1996,jones1996,wang1997,gutterres2002,amiot2002,bouloufa2009}. Other processes can take place while the molecule is in the excited state, like autoionization into molecular ions (observed for Na \cite{amelink2000}) or predissociation into a pair of one ground state atom and one excited atom (observed for Na \cite{molenaar1996} and K \cite{wang1998}). Highly excited molecular states can be studied by optical-optical double-resonance spectroscopy \cite{wang1997}. Long-range interactions between ground state atoms can also be probed by two-color PA spectroscopy according to a "$\Lambda$"-type level scheme. Accurate determination of the sign and magnitude of the scattering length $a$ can be obtained. This is possible either by precisely determining the position of the last vibrational levels in the ground state potential or by studying the intensity modulation of the PA spectrum \cite{drag2000a}.  Extensive reviews on various aspects of PA can be found in references~\cite{weiner1999,stwalley1999,jones2006}.

As we will review hereafter, PA performed in a cold and dense sample of laser-cooled atoms represents an efficient way to produce stable cold molecules, after the spontaneous decay of the photoassociated molecules. It is worthwhile to note that this possibility is not obvious at first glance. Indeed, cold atoms remain most of the time far from each other, so that PA most often produces excited molecules in long-range electronic states varying as $R^{-3}$ for homonuclear pairs. In contrast, electronic states linking two ground state atoms have a much shorter range determined by their van der Waals interaction varying as $R^{-6}$. Due to the Franck-Condon (FC) principle, the spontaneous decay of the excited molecules favors the formation of a pair of free atoms, which will escape from the trap and induce a change in the number of trapped atoms (trap-loss signal). Nevertheless cold ground state molecules can be produced quite efficiently through PA in some special cases, when the probability density of the excited molecular state can be increased at short internuclear distances.

\begin{figure}[htb]
\begin{center}
\includegraphics[width=10cm,height=15cm]{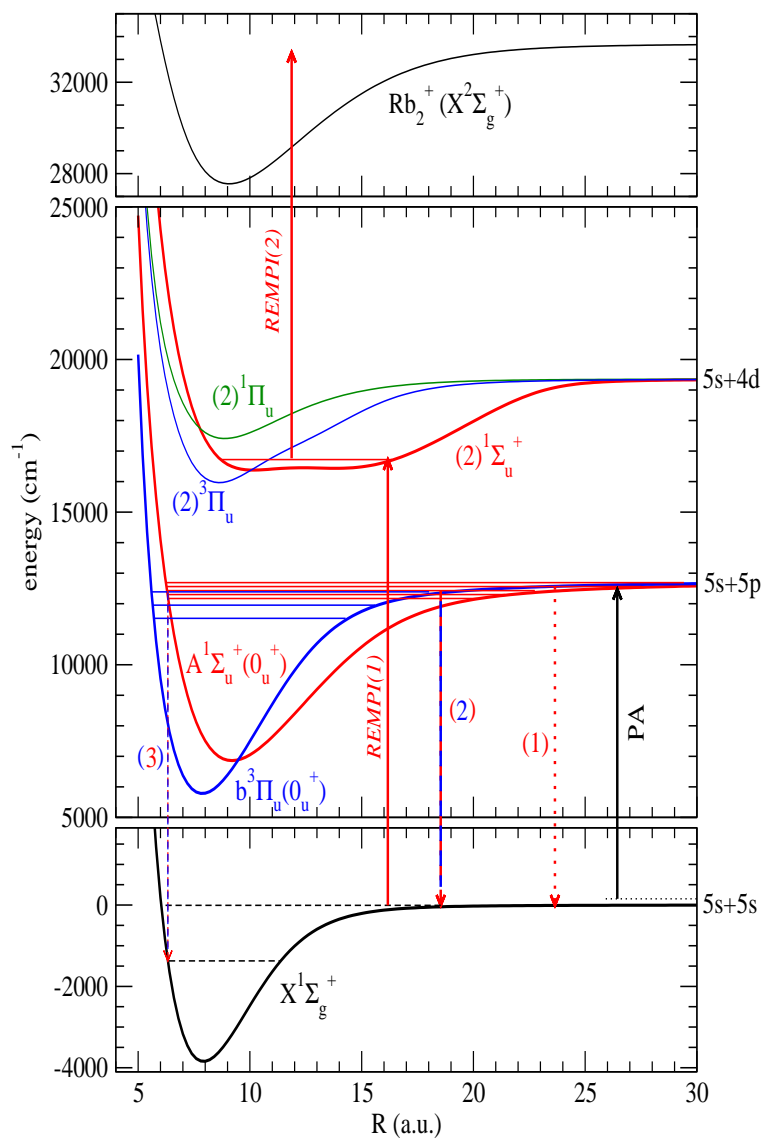}
\end{center}
\caption{Rb$_{2}$ molecular potential curves involved in one of the observed cold molecule formation processes, utilizing resonant coupling \cite{fioretti2007}, and in their detection. All potentials are drawn at the same scale, while levels in the excited states, not drawn on scale, illustrate the different level density in the $A^{1}\Sigma _{u}^{+}(0_{u}^{+})$ and $b^{3}\Pi _{u}(0_{u}^{+})$. Arrows correspond to spontaneous emission starting from (1) the external turning point of the PA level down to the uppermost vibrational level of the $X$ state, (2) from the intermediate turning point of the PA level induced by the resonant coupling down to moderately bound vibrational levels of the $X$ state, (3) from the inner turning point of the PA level induced by the resonant coupling down to strongly bound vibrational levels of the $X$ state. The other \textit{ungerade} electronic states correlated to the $5s+4d$ limit are also displayed for completeness.}
\label{fig:rb2}
\end{figure}

\begin{figure}[htb]
\begin{center}
\includegraphics[width=15cm,height=8cm]{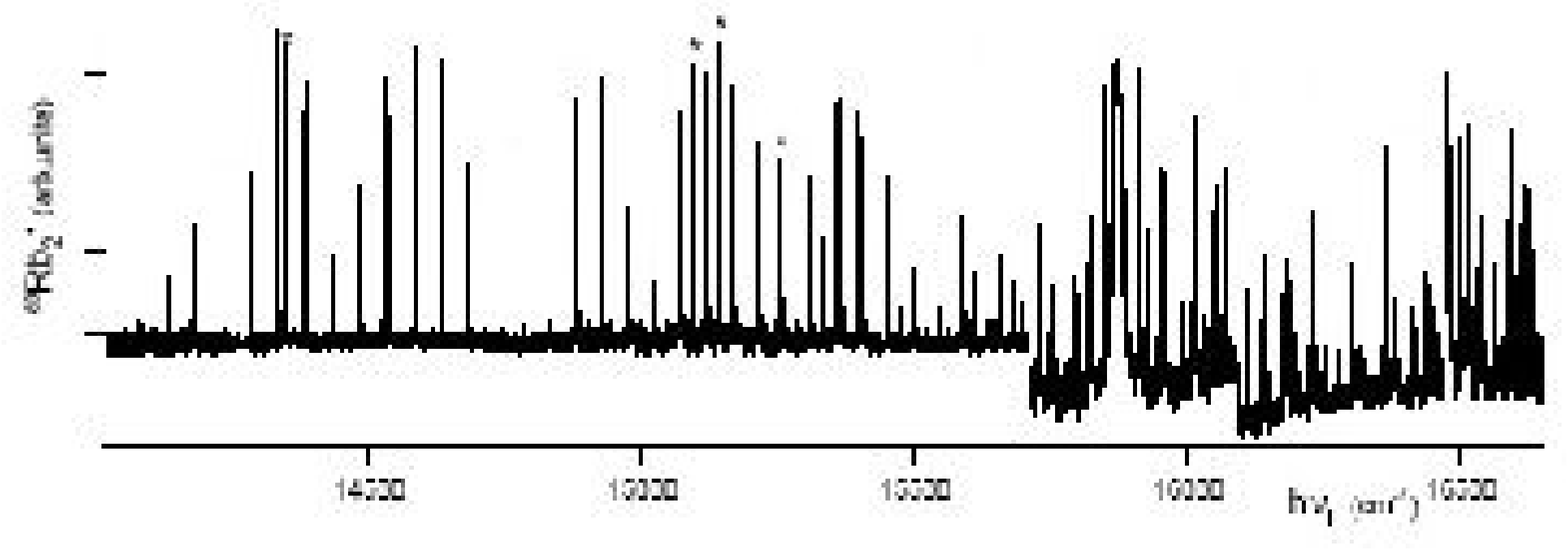}
\end{center}
\caption{Experimental \textit{REMPI} spectrum of $^{85}$Rb$_{2}$ dimers produced by PA with detuning $\delta_{PA}=-69$~cm$^{-1}$ from the $(5S+5P_{1/2})$ asymptote. The shift of the baseline is due to the change of the dye used for the ionization laser. The asterisks indicate ghost lines produced by two-photon transitions to atomic Rydberg states. Reprinted with permission from Lozeille {\it et al} \cite{lozeille2006}.}
\label{fig:detect}
\end{figure}

The key point for their observation has been to set up an appropriate detection technique, imposed by the low density of the formed molecules. A good choice was to selectively photoionize the molecules with a Resonantly Enhanced Multiphoton Ionization (REMPI) scheme \cite{johnson1981}, and detect them after a mass selection by an ion detector (channeltron or microchannel plates). In general it is sufficient to let the particles travel along a few centimeter path to separate atomic ions from molecular ions by time-of-flight (TOF). However, when the masses of the products are very close to each other (like LiCs$^+$ and Cs$^+$ \cite{kraft2006}), a more sophisticated TOF apparatus, like a Wiley-McLaren detector, should be used. The REMPI method was first used in a cold molecule experiment by Fioretti {\it et al}~\cite{fioretti1998}, and applied afterwards in most experiments devoted to the formation of cold dimers by PA. An intense nanosecond laser, pulsed at fairly low rate ($\simeq 10-20$~Hz) is in general preferable to a c.w. laser chopped at high rate, as high ionization rates preferentially detect excited state molecules, whose production rate is much larger than for stable dimers. On a longer timescale, the number of stable molecules becomes larger than the number of excited molecules, because they are accumulated in the ionization region, due to their low velocity. An example of formation and detection processes is shown in Fig.~\ref{fig:rb2}, while a REMPI spectrum is shown in Fig.~\ref{fig:detect}. Several molecular intermediate states can be used in the REMPI process. For example a convenient choice is the $(2)^{3}\Pi_{g}$ state in Cs$_2$ or Rb$_2$, which is responsible of the so-called diffuse bands in absorption spectra of high-temperature alkali vapors~\cite{pichler1983}. This state offers good FC factors for transition with almost all vibrational states of the lowest triplet state, and is located at an energy lower than the two-photon ionization threshold of the atomic ground state, which makes the molecular ionization much more probable than the atomic one.

The formed molecules have almost the same translational temperature than the starting atomic sample, as the recoil energy associated to the photon absorption and emission is negligible. Only the lowest rotational levels are populated as the centrifugal barrier prevents the initial atom pairs to collide in partial waves higher than the lowest ones ($s$, $p$, $d$). The stable molecules are typically formed in high vibrational levels of the singlet ground state or of the lowest triplet state. In principle their population can be transferred down to lower vibrational levels by optical pumping processes, where the spontaneous decay step of the formation process is  replaced by a stimulated emission step. In this case the process can be coherently driven with the advantage of populating a single final state. In real experiments, as it will be shown in the following sections, the process has limitations due to losses and decoherence. Another possibility is the stimulated Raman adiabatic passage (STIRAP) process made by a counter-intuitive pulse sequence, where the first pulse couples the molecular bound levels and the second one makes the coupling with the continuum \cite{mackie2000}. An alternative proposal for photoassociation is the utilization of fast pulses at high repetition rate with pulse duration ranging from hundreds of ns \cite{shapiro2007} down to tens of fs \cite{vala2001,koch2006a}. By appropriately chirping the pulses it should be possible to efficiently couple the starting continuum wave function to the bound state one, ensuring adiabatic following conditions. Coherent control techniques like feedback-learning loops could be employed. Raman or short pulses can be also used to transfer the population of cold molecules from high to low vibrational levels \cite{koch2006,peer2007}. Preliminary experiments have been performed using chirped femtosecond laser pulses to make PA \cite{brown2006,salzmann2006}. These experiments showed a quenching of the molecular signal rather than an enhancement; however a signature of the coherence of the process was found, giving an indication that coherent control of PA could be accomplished. Very recent experiments \cite{salzmann2009} observed the transient dynamics of the PA of cold Rb atoms using shaped fs laser pulses. Time-resolved pump-probe spectra revealed coherent oscillations of the molecular formation rate, assigned to coherent transient dynamics in the electronic excitation.



\begin{figure}[htb]
\begin{center}
\includegraphics{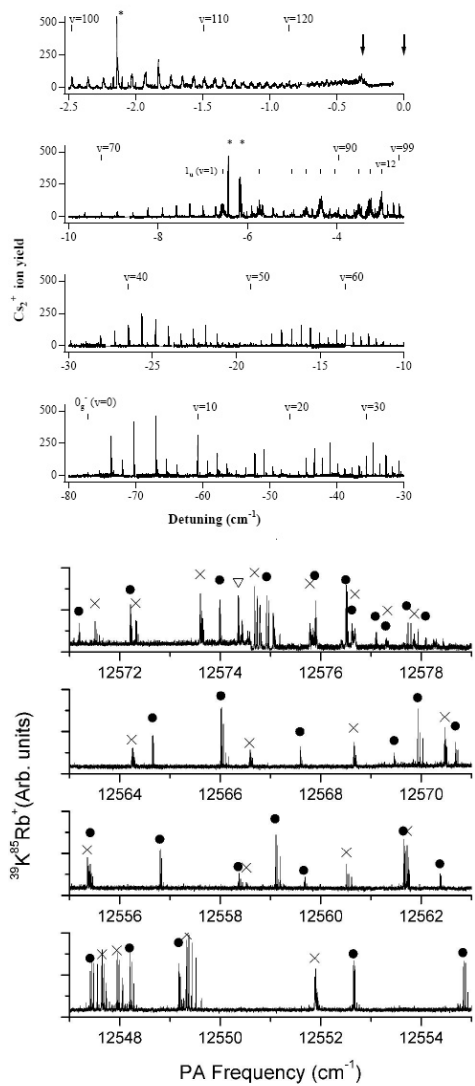}
\end{center}
\caption{Photoassociative spectra of Cs$_2$ molecules (upper part) and KRb molecules (lower part), obtained after REMPI and molecular ion detection. For Cs$_2$ PA is done below the $6S_{1/2}+6P_{3/2}$ asymptote, while for KRb PA is done below $K(4S_{1/2})+ Rb(5P_{1/2})$ asymptote. Note the difference in the level density between the two spectra, due to the different shape of the excited state potentials near the asymptotes (varying as 1/R$^3$ vs. 1/R$^6$, respectively). For Cs$_2$, the transitions are due to the $0_{g}^{-}$ and $1_{u}$ states, while for KRb the $\Omega=0,1,2$ bands are indicated by circles, crosses and triangles respectively. Reprinted with permission from Fioretti {\it et al} \cite{fioretti1999} and from Wang {\it et al} \cite{wang2004}.}
\label{fig:spectrum}
\end{figure}

The first experimental observation of cold molecules produced through PA, has been achieved by the group at Laboratoire Aim\'e Cotton (Orsay, France), where Cs$_{2}$ molecules falling out of a MOT have been observed ~\cite{fioretti1998}. Cold molecules were produced in the  lowest (metastable) triplet state $a^{3}\Sigma _{u}^{+}$  as the decay products of the $0_{g}^{-}$ pure long-range excited state. The specific reaction for Cs reads:

\begin{equation}
Cs(6s)+Cs(6s)+h\nu _{PA}\rightarrow Cs_{2}^{\ast}(0_{g}^{-}(6S+6P_{3/2};v,J))\rightarrow Cs_{2}(a^{3}\Sigma
_{u}^{+};v^{\prime },J^{\prime })+h\nu ^{\prime}
\end{equation}

where the $0_{g}^{-}$ state is one of the four (among $0_{g}^{-}$, $0_{u}^{+}$, $1_{u}$, $1_{g}$) attractive electronic states connected to the $6S+6P_{3/2}$ asymptote, which can be optically excited from the ground state (see Fig.\ref{fig:longrangewells}. This is one of the specific cases where the spontaneous decay rate into stable molecules is enhanced. It happens when PA populates levels corresponding to 'pure long-range molecules'~\cite{stwalley1978}, in which the entire vibrational motion takes place between intermediate ($\approx 15a_0$) and large (well beyond 25$a_0$) internuclear distances. This feature induces a "$R$-transfer" of the probability density from large distances towards short distances. The inner turning point of such double-well states represents a favorable Condon point for decaying into a stable molecule, especially for the $0_{g}^{-}$($nS+nP_{3/2}$) states of cesium ($n=6$) and rubidium ($n=5$). The production rate of cold molecules can reach values up to a few millions of molecules per second for Cs~\cite{drag2000}). The cold molecules formation through PA into pure long-range molecules works excellently in cesium, through both $0_{g}^{-}$ and $1_{u}$ long-range states ~\cite{fioretti1998,comparat1999} (see Figure~\ref{fig:spectrum}), and also in rubidium through the $0_{g}^{-}$ long-range state ~\cite{gabbanini2000,fioretti2001}. The mechanism becomes less favorable for the lighter alkalis, because the long-range wells are located at larger distances. A similar mechanism has been proposed \cite{pichler2004b} - but not yet observed- to form deeply-bound ground state Cs$_2$ molecules, relying on the excitation of the highly-excited $(3)^`\Sigma_u^+ (6S+7S)$ state. Its potential curve has two wells separated by a barrier located around 13$a_0$ induced by the interaction between the covalent and the ion-pair configurations within the molecular state. It favors a "$R$-transfer" if the PA excitation is appropriately chosen for reaching the top of the hump.

Another efficient process leading to the formation of cold molecules relies on the so-called {\it resonant coupling} in Rb$_2$ and Cs$_2$: the strong spin-orbit coupling between the $A^1\Sigma_u^+(nS+nP)$ and $b^3\Pi_u(nS+nP)$ states induces resonant interaction between vibrational levels of the resulting $0_u^+$ states \cite{dion2001}. Like in the previous case, it creates a favorable FC point at short distances (a "$R$-transfer") suitable for direct spontaneous emission down to vibrational levels of the singlet ground state of Cs$_2$ \cite{dion2001,kokoouline2002,pichler2006,jelassi2008b} and Rb$_2$ \cite{bergeman2006,huang2006,jelassi2006b,fioretti2007,pechkis2007}.

Ground-state potassium molecules in deeply bound levels of the singlet $X^{1}\Sigma_{g}^{+}$ ground state have been observed after the direct spontaneous decay of the $A ^{1}\Sigma_{u}^{+}(4S+4P)$ state ~\cite{nikolov1999}. The molecule formation rate is quite low (of the order of 10$^{3}$ molecules per second), as the emission mainly takes place at the inner turning point of a standard molecular potential well with a steep wall, i.e. where the probability density is low. A more efficient production of ground state K$_2$ molecules through a two-step pumping scheme via highly-excited electronic (Rydberg) states like the $(5)^1\Pi_u(4S+4D)$ or the $(6)^1\Pi_u(4S+4F)$ has been demonstrated~\cite{nikolov2000}. It again relies on a "$R$-transfer" mechanism towards short-range potential wells of the Rydberg states, which favors the production of K$_2$ molecules in deeply-bound levels.

Cold sodium dimers in the last two vibrational levels ($v=$14,15) of the lowest triplet state have been detected \cite{fatemi2002} after PA into the $0_{g}^{-}(3S+3P_{1/2}$) state. In this experiment the molecular detection was done by resonant ionization using a narrow band c.w. laser, with a resolution limited just by the radiative lifetime of the intermediate state in the ionization path.

\begin{figure}[htb]
\begin{center}
\includegraphics{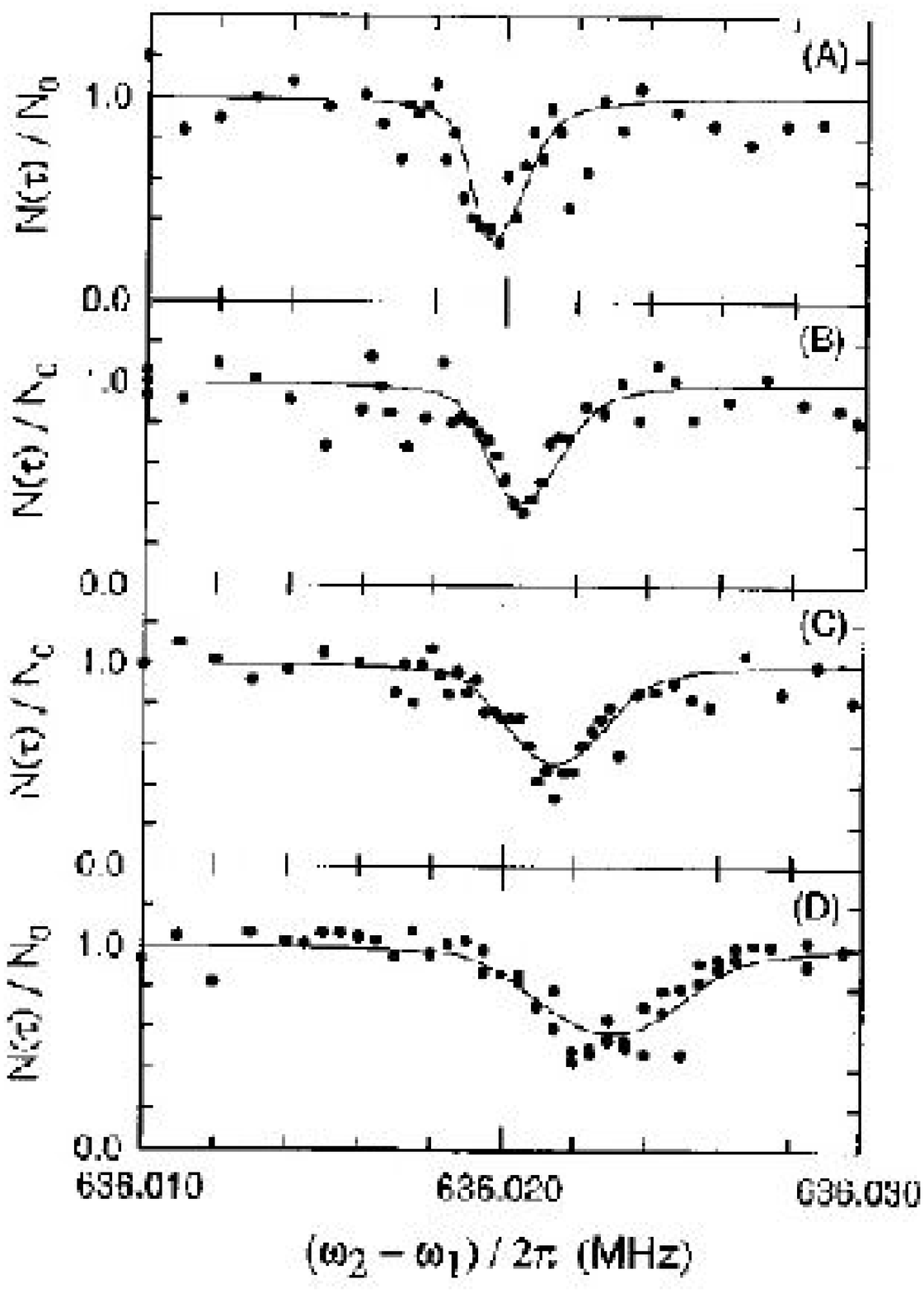}
\end{center}
\caption{Stimulated Raman line shapes in a Rb BEC for four different peak condensate densities: (A) n=0.77 10$^{14}$ cm$^{-3}$; (B) n=1.22 10$^{14}$ cm$^{-3}$; (C) n=1.75 10$^{14}$ cm$^{-3}$; and (D) n=2.60 10$^{14}$ cm$^{-3}$. Each spectrum shows the fraction of atoms remaining in the condensate after illumination by the two coherent laser fields, as a function of the laser frequency difference, due to molecule formation.
The increase in width and center frequency of the resonance with density arises from the atom-condensate and molecule-condensate mean-field interactions. Reprinted with permission from Wynar {\it et al} \cite{wynar2000}.}
\label{fig:wynar}
\end{figure}

Rubidium molecules have been produced in a condensate of $^{87}$Rb atoms by a stimulated Raman process \cite{wynar2000}. A first laser is tuned to the free-bound transition to a rovibrational level of the excited $0_{g}^{-}(5S+5P_{1/2}$) bound state, with a small detuning necessary to minimize population of the excited state. A second laser beam closes the Raman process to the second-to-last vibrational level of the lowest triplet state. The Raman process is observed through the atom loss from the condensate. The width of the transition is extremely narrow (a few KHz), showing an increase as a function of the peak condensate density due to mean-field interactions (Fig.~\ref{fig:wynar}). A shift of the line has also been observed. Besides the exact energy position of the bound level, the analysis of the line shape yielded the atom-molecule collisional parameters.

Soon after the first experiments on homonuclear dimers, a large activity has been devoted to the formation of ultracold heteronuclear molecules composed of two different alkali atoms. As it is discussed later in this review, such molecules exhibit a permanent electric dipole moment which induces long-range interactions among them, opening new possibilities for many applications. In contrast with a pair of identical alkali atoms, an alkali atom in its ground state interacts with an alkali atom of a different species at large distances through the van der Waals interaction varying as $\propto$R$^{-6}$. As a consequence, the spatial overlap (FC factors) between the initial collisional wave function and the vibrational wave functions of the excited molecular states are significantly smaller than for the homonuclear case. However the FC factors for spontaneous decay back into ground state molecules is enhanced due to the similar shape of the potential curves in the ground and excited states. The resulting overall efficiency for producing ground-state molecules becomes therefore comparable to homonuclear dimers \cite{azizi2004}.

Ground state RbCs molecules have been produced in the lowest triplet state $a^{3}\Sigma^{+}$ by PA in a dual-species MOT \cite{kerman2004a}. A photoassociation laser tuned below the lowest excited asymptote Rb($5S$)+Cs($6P_{1/2}$), created RbCs molecules in the excited $0^{-}$ state resulting from the spin-orbit coupling of the $(2) ^{3}\Sigma^{+}$ and $b ^{3}\Pi$ states \cite{kerman2004}. The decay process produced RbCs in the lowest triplet state with a rate as high as $5 \times 10^{5}$~s$^{-1}$ for the most populated vibrational level ($v=$37). They were detected by state selective REMPI through the intermediate $(2)^{3}\Sigma^{+}$ and $B^{1}\Pi$ coupled states correlated to Rb($5S$)+Cs($6P$).

Observation of cold KRb molecules has been first reported in a double species MOT, presumably produced by PA from the trapping lasers and spontaneous emission \cite{mancini2004}. In another experiment, KRb dimers in the lowest triplet state and in the singlet ground state have been produced and detected by REMPI in a two-species dark-SPOT MOT by a PA laser tuned below the K($4S$)+Rb($5P_{1/2}$) asymptote followed by spontaneous decay \cite{wang2004}. A PA spectrum of KRb molecules is shown in Figure~\ref{fig:spectrum}. All eight attractive potential curves converging to the K($4S$)+Rb($5P_{1/2}$) and K($4S$)+Rb($5P_{3/2}$) asymptotes, i.e. two $\Omega=0^{+}$, two $\Omega=0^{-}$, three $\Omega=1$ states and one $\Omega=2$ state, have been identified and classified in the PA spectrum of the detected products \cite{wang2004a}. State-selective detection by REMPI has been analyzed using  the $4 ^{1}\Sigma^{+}$, $5^{1}\Sigma^{+}$, $4 ^{3}\Sigma^{+}$ and $3 ^{3}\Pi$ states as resonant intermediate states \cite{wang2005}. The analysis has allowed the assignment of vibrational levels bound by an energy as large as 30~cm$^{-1}$ of both the $a^{3}\Sigma^{+}$ and $X^{1}\Sigma^{+}$ states. The detection technique has been further refined  using the depletion spectroscopy technique, that consisted in exciting the produced stable molecules with an additional c.w. laser. In this way a dip in the ion signal is produced when the depletion laser is resonant, yielding a resolution allowing the assignment of the rotational structure \cite{wang2007}. Other experiments observed the formation of cold heteronuclear dimers of NaCs \cite{haimberger2004,haimberger2006} and LiCs \cite{kraft2006}. Further alkali pairs are currently under investigation \cite{li2008d,marzok2009}.

\subsubsection{Molecules formed after magnetoassociation of cold alkali atom pairs}
\label{sssec:magnetoass}

After the achievement of Bose-Einstein condensation of alkali atoms \cite{anderson1995,bradley1995,davis1995} magnetically-tuned Feshbach resonances have been the subject of a growing number of studies, which are reviewed for instance in ref.\cite{kohler2006}. Briefly, a Feshbach resonance occurs when, in the course of a scattering process, a bound state of a closed channel is coupled through some interaction with the scattering flux in the entrance channel. In ultracold alkali atom collisions, this occurs within the hyperfine manifold of states, when a bound state of a closed hyperfine channel is located at almost zero energy, corresponding to the entrance channel. In a magnetic field the resonance energies depend on the field strength through the Zeeman effect of the different hyperfine levels. The zero-energy resonance position is determined by the field strength at which the energy of a diatomic vibrational bound state becomes degenerate with the threshold for dissociation into an atomic pair at rest. The $s$-wave scattering length has a simple expression near the resonance given by:
$a(B)=a_{bg}(1-\frac{\Delta}{B-B_{0}})$, where $a_{bg}$ is the background scattering length, $\Delta$ is the resonance width and B$_0$ is the Feshbach resonance position. For positive scattering lengths, the state describes a stable molecule in the absence of collisions with the background gas. The value of the scattering length can be tuned by varying the position of Feshbach resonances with an external magnetic field \cite{tiesinga1993}, and can be used for the conversion of atom pairs into molecules \cite{timmermans1999}. Let us note that Feshbach resonances can also be induced by optical means using a single laser \cite{fatemi2000,theis2004,enomoto2008a}, or a stimulated Raman scheme through a bound molecular state \cite{thalhammer2005}.

The applications of Feshbach resonances span from the Bose-Einstein condensation of $^{85}$Rb \cite{cornish2000} and Cs \cite{weber2003} to studies of the collapse of condensates with negative scattering lengths \cite{donley2001}. Application of Feshbach resonances for ultracold molecule formation has first been predicted in ref.\cite{abeelen1999} and demonstrated later in ref.\cite{donley2002}. The coherence properties of the atomic BEC are transferred to the molecules, as demonstrated by two-photon Bragg scattering, where diffracted molecules recoil with the same momenta as the atoms but expand with half the velocity due to their mass \cite{abo-shaeer2005}. After the first observation, several groups have concentrated experimental effort to produce so-called Feshbach molecules. Specific reviews on molecule formation through magnetic Feshbach resonances can be found in \cite{kohler2006,hutson2006}. For these experiments new schemes for the detection of Feshbach molecules have been developed. These techniques involve radio frequency (rf) photodissociation, atom loss and recovery, as well as the spatial separation of molecules from the remnant atomic cloud followed by their dissociation using magnetic field sweeps. The separation of Feshbach molecules from the atomic gas is achieved via the Stern-Gerlach approach \cite{herbig2003}, probing the magnetic moments of dimers at magnetic fields away from the zero energy resonance. Weakly-bound heteronuclear molecules produced near a Feshbach resonance can be directly imaged by light resonant with atoms, thanks to the $1/R^6$ dependence of the potentials in both ground and excited states \cite{zirbel2008}.

\begin{figure}[htb]
\begin{center}
\includegraphics[width=10cm,height=8cm]{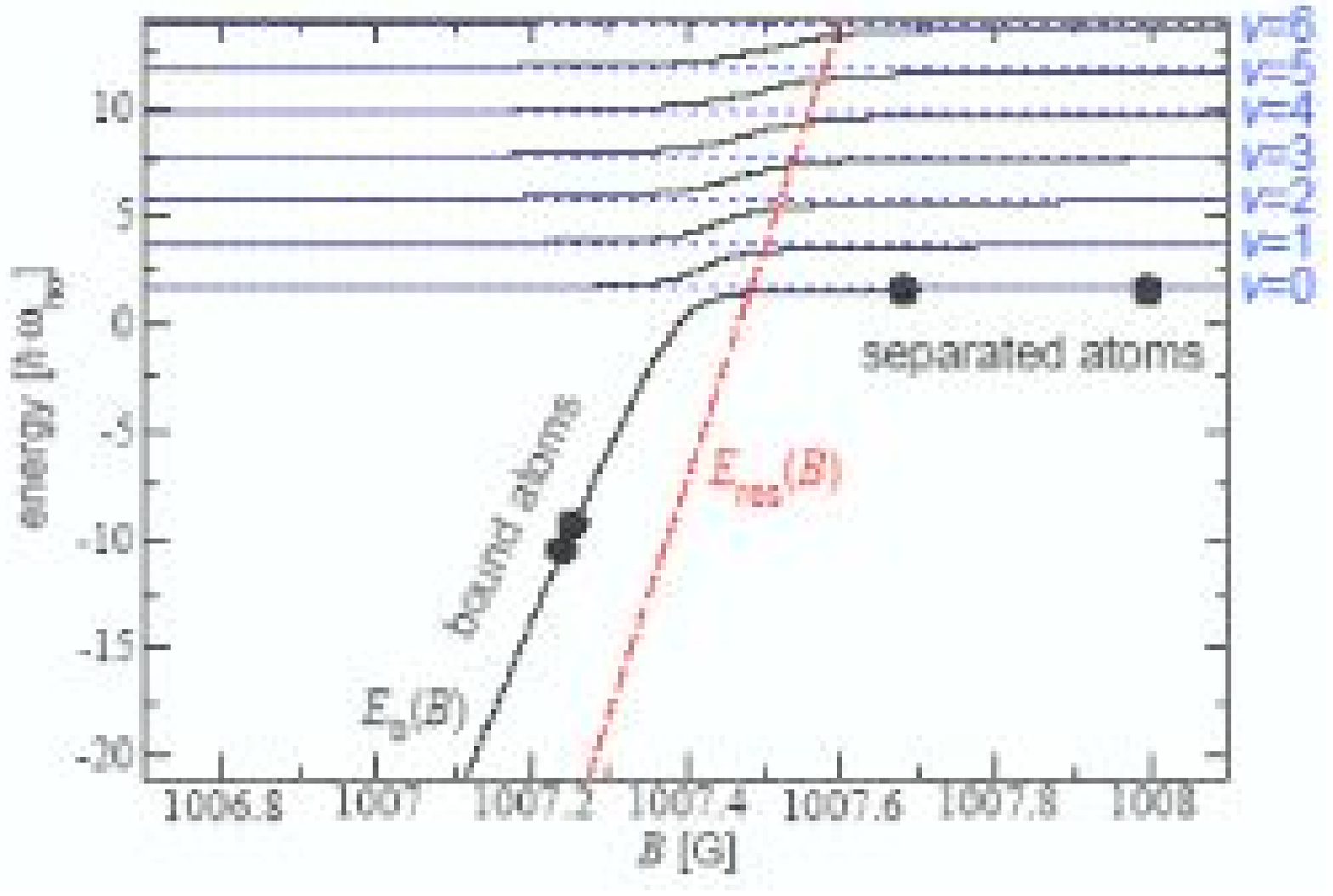}
\end{center}
\caption{Scheme of molecular association of a pair of ground state $^{87}$Rb atoms via a downward magnetic field sweep in a spherical harmonic atom trap. The bare vibrational levels ($v = 0-6$) associated with the background scattering quasi-continuum (i.e. the discretized levels of the trap with positive energy) and the Feshbach resonance energy, E$_{res}$(B), are indicated by dotted and dashed lines, respectively. Solid curves refer to the magnetic field dependence of the dressed energy levels near the zero- energy resonance position at 1007.4~G. Reprinted with permission from K\"{o}hler {\it et al} \cite{kohler2006}.}
\label{fig:kohler}
\end{figure}

The first experiment on Feshbach molecules has been achieved a BEC of bosonic $^{85}$Rb atoms exposed to pairs of magnetic field pulses in the vicinity of the 155~G zero-energy resonance \cite{donley2002}. This experiment probed the regime of strong interactions where the magnitude of the scattering length is comparable to the average interatomic distance. Such perturbations induces three distinct components of the gas. The oscillatory behavior of their relative proportions as a function of the time delay between the pulses implied an interpretation in terms of Ramsey interference fringes due to a superposition state of separated atoms and Feshbach molecules. Subsequent experiments improved the production efficiency by using  magnetic field sweeps from negative to positive scattering lengths across a zero-energy resonance (Figure~\ref{fig:kohler}). If the magnetic sweep is sufficiently slow, the atomic energy adiabatically follows the magnetic field change. The ramp speed determines the conversion efficiency, which vanishes for very fast ramps. A reverse sweep dissociates the molecules back into atoms, with a relative velocity settled by the ramp speed \cite{durr2004}. The technique of magnetic sweep has been applied to atomic Bose-Einstein condensates as well as to ultracold gases of fermions. The conversion efficiency depends on the sweep rate and on the phase space density, while it does not depend on the statistics, as demonstrated in \cite{hodby2005}, where a simple model reproduced the conversion efficiency of both bosons and fermions. Another technique consists in modulating the field at a frequency corresponding to the energy difference between the free atomic channel and the molecular bound state \cite{thompson2005}.

Molecular formation using magnetic field sweeps across Feshbach resonances has been first observed with bosonic species like Cs\cite{herbig2003}, Na \cite{xu2003} and $^{87}$Rb \cite{durr2004}. Heteronuclear bosonic molecules have also been produced using interspecies Feshbach resonances, as demonstrated for $^{85}$Rb$^{87}$Rb dimers \cite{papp2006} and $^{41}$K$^{87}$Rb molecules \cite{weber2008}. In homonuclear samples, experiments started from atoms in optical traps, pumped to the lower energy hyperfine state, which is stable against collisions. A magnetic field sweep converts a fraction of the atoms into dimers. The dimers are produced in very high excited levels with binding energies smaller than 0.1~cm$^{-1}$ and a wave function extending over several thousands of Bohr radii. Although in some cases, as in ref.\cite{xu2003}, molecules were generated in the quantum-degenerate regime, they were not referred to as a molecular Bose-Einstein condensate, because their lifetime was not sufficient to reach full thermal equilibrium. The lifetime of the molecules turned out to be indeed very short, becoming slightly longer (from 1 to tens of ms) near resonance. Inelastic atom-dimer collisions like the vibrational quenching collisions can explain such lifetimes, as the inelastic rate constants $K_{AD}$ are of the order of 10$^{-10}$~cm$^{3}$~s$^{-1}$. The molecules are heated by these collisions and ejected from the trap. Atom-molecule collisions can be eliminated by 'purification', i.e. by a resonant pulse or another process that removes atoms from the sample. In presence of an optical lattice for low occupation number, molecule-molecule collisions can be avoided, and the molecular lifetime can be increased by a large factor (see the next subsection).

Metastable Cs$_2$ molecules in high rotational states with a lifetime larger than 1~s have been recently produced \cite{knoop2008}. The molecules are created by Feshbach association in a $g$-wave state and transferred via an intermediate state to an $l$-wave ($\ell$=8, where $\ell$ is the rotational angular momentum of the atom pair) state. The molecules in this state are stable even at magnetic fields corresponding to energies above the dissociation threshold due to the large centrifugal barrier and the weak coupling with other states.

Other experiments investigated the molecular formation in fermionic species. Collisions between two fermions in the same hyperfine state are suppressed by the Pauli exclusion principle at ultralow temperatures. Indeed, $s$-wave collisions are forbidden because of the antisymmetry of their total wave function and therefore , two spin-component mixtures must be prepared to achieve low temperature collisions and evaporative cooling. Two-spin-component mixtures of fermions are in general cooled in optical traps (although $^{40}$K two-state mixtures can be evaporatively cooled also in a magnetic trap).

\begin{figure}[htb]
\begin{center}
\includegraphics[width=13cm,height=10cm]{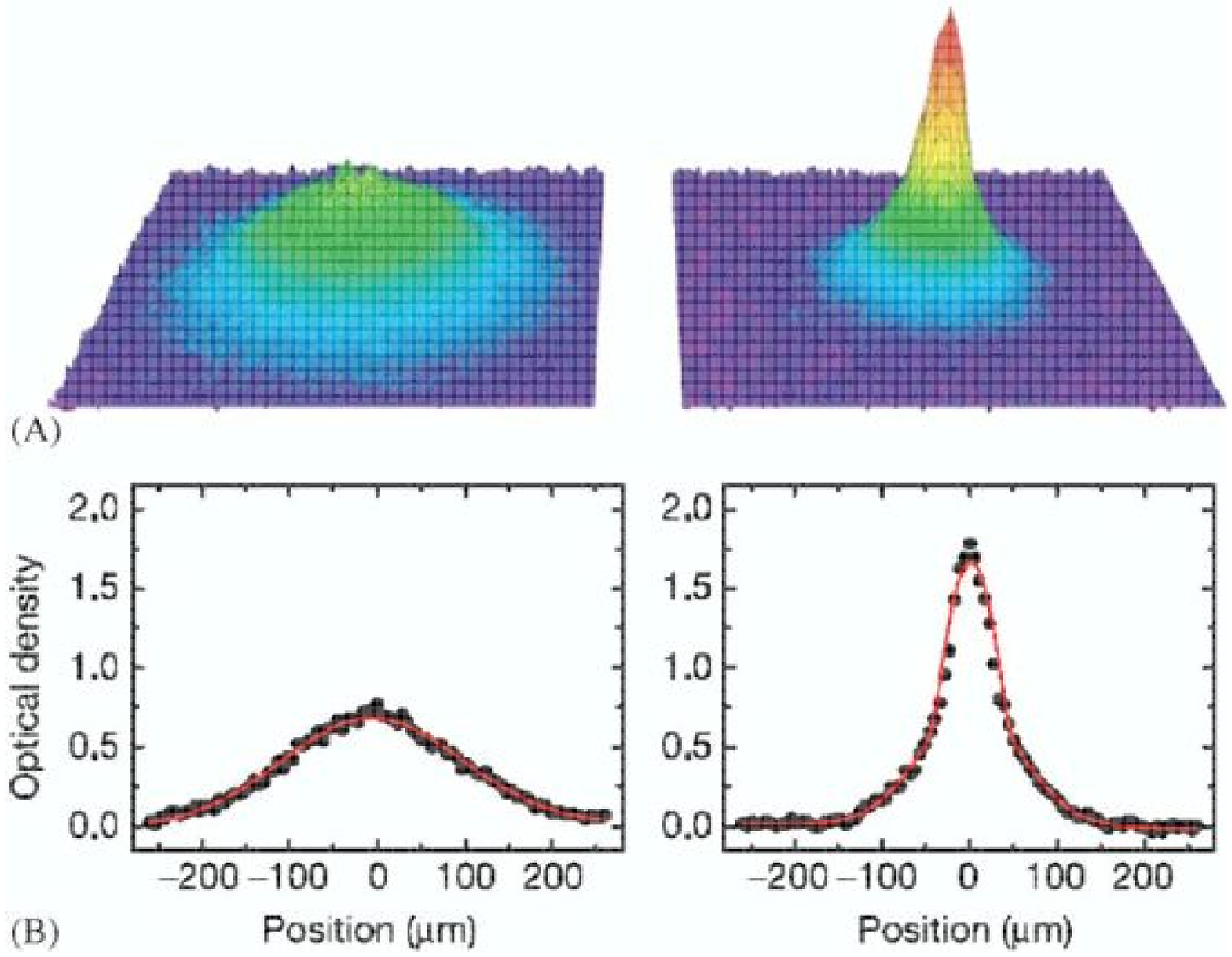}
\end{center}
\caption{(A): images of a molecular cloud of $^{40}K_2$ after 20 ms of free expansion, above the BEC critical temperature (left) and below it (right), showing the tight spatial peak characteristic of a condensate. (B): corresponding 1-D distribution of the optical density of the molecular cloud, fitted by a combination of a Gaussian and a Thomas-Fermi distribution. Reprinted with permission from Greiner {\it et al} \cite{greiner2003}.}
\label{fig:greiner}
\end{figure}

The mixture can be prepared from a single state by radiofrequency (rf) spectroscopy; decoherence mechanisms help to transform the prepared coherent state into an incoherent mixture. In the first experiment looking for Feshbach molecules formed from fermionic atoms ($^{40}$K), the lifetime of the dimers created through magnetic sweep turned out to be short ($\approx$1 ms) \cite{regal2003}. Experiments were also performed on fermionic $^{6}$Li atoms, both by sweeping the magnetic field across a Feshbach resonance \cite{cubizolles2003,strecker2003} and by using the enhancement of three-body collision rate near a Feshbach resonance to efficiently form the dimers \cite{jochim2003}. In these cases the lifetime of the produced $^{6}$Li$_2$ molecules was found very long, larger than 1~s, and even larger near the resonance where the scattering length is large and positive. This was later confirmed for $^{40}$K \cite{regal2004}. The explanation of such long lifetimes comes from the Fermi statistics that acts by suppressing atom-dimer and dimer-dimer collisions at very long range \cite{petrov2004}. Long lifetimes (beyond 100~ms) have been recently measured for heteronuclear $^{6}$Li$^{40}$K molecules created from fermionic species \cite{voigt2009}.

Soon after the reports about the observation of the huge lifetime of the fermionic dimers, experiments on $^{6}$Li and $^{40}$K showed the formation of Bose-Einstein condensate of Li$_{2}$ \cite{jochim2003a,zwierlein2003} and K$_{2}$ \cite{greiner2003}. Li$_{2}$ dimers have been condensed using three-body recombination near a Feshbach resonance while  the magnetic sweep technique was used for K$_{2}$. The onset of condensation was observed, as shown in Figure~\ref{fig:greiner}, by the appearance of a density peak below a critical temperature after reconverting the molecules into atoms.

\subsubsection{Formation of molecules in optical lattices}
\label{sssec:lattices}

The use of a three-dimensional optical lattice to increase the efficiency of molecular formation through PA thanks to the tight atomic confinement was suggested in \cite{jaksch2002}. In a first experimental realization \cite{rom2004}, a Bose-Einstein condensate of $^{87}$Rb atoms was loaded in a 3D optical lattice and converted into a Mott insulator by increasing the potential depth. In the Mott insulator phase, atoms are localized at individual lattice sites with shells having an exact number of atoms per site (in this case single atoms or atom pairs) and the atomic motion through the lattice is blocked due to the repulsive interactions between the atoms \cite{greiner2002}. The lattice modifies the long-range part of the interatomic potential, changing the free-bound PA transition into a bound-bound one. A two-photon Raman transition converted the atom pairs into molecules in one of the two uppermost vibrational levels of the $a^{3}\Sigma _{u}^{+}$ state. The spectra showed a progression of resonances with a spacing due to the quantized motional states of the dimers in the lattice \cite{rom2004}. In a similar experiment, atoms in a Mott insulator were exposed to two pulses in a stimulated Raman scheme and coherent oscillations between an atomic and a molecular quantum gas were observed as a function of the PA pulse duration \cite{ryu2005}.
An optical lattice can also be used to increase the lifetime of molecules created from bosonic atoms. In \cite{thalhammer2006}, $^{87}$Rb$_2$ dimers were created through an adiabatic ramp through a Feshbach resonance in a 3D lattice. By using a combination of microwave and light pulses, the sample was purified from remaining atoms, preventing atom-molecule inelastic collisions. Loading individual molecules into the sites and using a sufficiently deep optical lattice, molecule-molecule collisions are suppressed, resulting into a long molecular lifetime (700~ms). The conversion efficiency from atoms to molecules turned out to be extremely high (95 \%). In a later experiment the loading of exactly one molecule per lattice site was demonstrated \cite{volz2006}.

\begin{figure}[htb]
\begin{center}
\includegraphics[width=10cm,height=10cm]{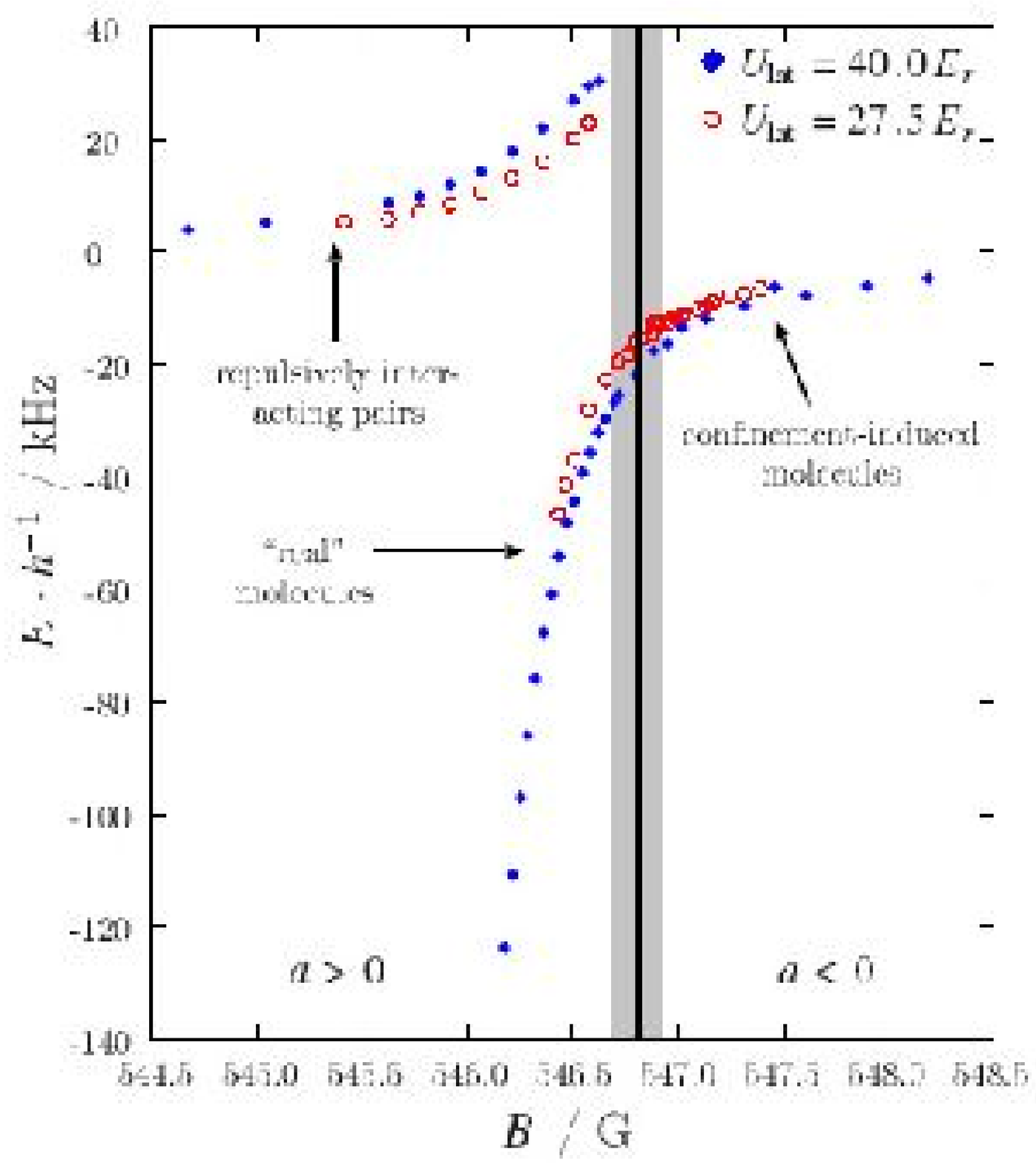}
\end{center}
\caption{Binding energy of heteronuclear $^{40}$K-$^{87}$Rb molecules in an optical lattice for two different lattice depths U$_{lat}$ in units of the $^{87}$Rb recoil energy. The center of the Feshbach resonance is located at 546.8(1) G. Attractively bound molecules, which are confinement-induced at a positive detuning with respect to the resonance center, and real molecules, which are stable in free space below the center of the resonance, are observed. In addition, repulsively interacting pairs with a positive binding energy below the resonance are also observed. Reprinted with permission from Ospelkaus {\it et al} \cite{ospelkaus2006}.}
\label{fig:ospelkaus}
\end{figure}

Molecules in optical lattices have also been produced from fermionic atoms \cite{stoeferle2006}. In this experiment the confinement of the molecules induced by the presence of the lattice potential, has been observed in the region of negative scattering length. More recently heteronuclear dimers have been created starting from a bosonic atomic species and a fermionic one \cite{ospelkaus2006}. Using rf-association near a Feshbach resonance between atoms in specific hyperfine levels $\left|F,m_F\right\rangle$ $^{40}$K$\left|9/2,-9/2\right\rangle$ and $^{87}$Rb$\left|1,1\right\rangle$ atoms, the formation of heteronuclear KRb dimers has been observed. A mixture of $^{40}$K$\left|9/2,-7/2\right\rangle$ and $^{87}$Rb$\left|1,1\right\rangle$ atoms has been prepared inside an optical lattice and analyzed by rf-spectroscopy. The appearance of a peak detuned with respect to the $^{40}$K$\left|9/2,-7/2\right\rangle\rightarrow^{40}$K$\left|9/2,-9/2\right\rangle$ atomic peak has been assigned to the molecular formation. By varying the magnetic field value with respect to the Feshbach resonance center, real molecules, confinement-induced molecules or repulsively-bound atom pairs \cite{winkler2006} were observed, as shown in Fig.~\ref{fig:ospelkaus}.

\subsubsection{Creating cold molecules in their lowest vibrational level}
\label{sssec:v=0}

Both association techniques presented so far have the drawback of producing molecules in high-lying vibrational states. For stability reasons and for many applications it is important to create dimers in the lowest rovibronic state. Some recent experiments obtained spectacular results on this aspect.

\begin{figure}[htb]
\begin{center}
\includegraphics[width=10cm,height=15cm]{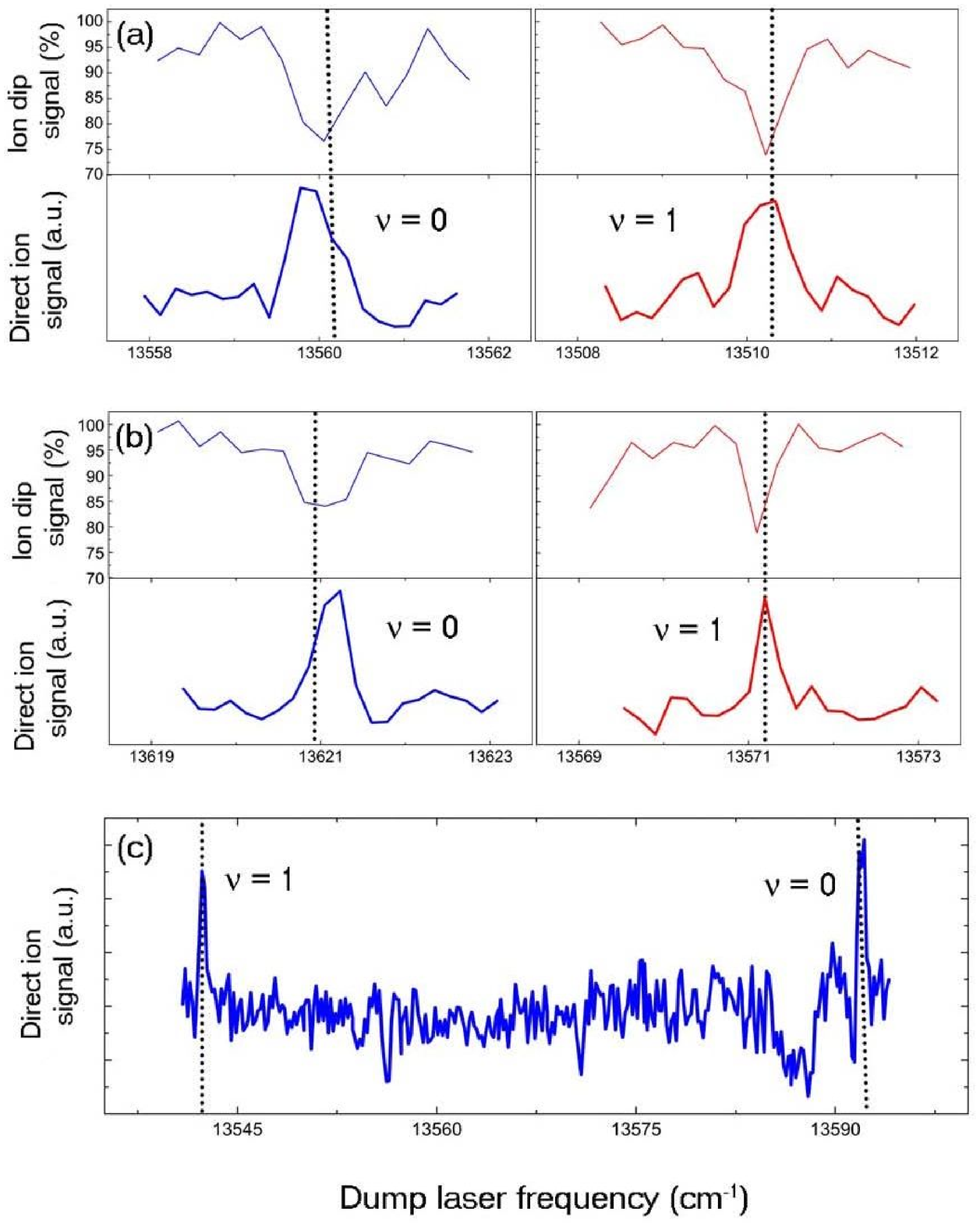}
\end{center}
\caption{Observation of $X^1\Sigma^+$ v=0,1 state RbCs molecules. Results are shown for excited state depletion (upper) and direct detection (lower) for three consecutive excited states, located at energies of (a) 9754.26 cm$^{-1}$, (b) 9814.60 cm$^{-1}$, and (c) 9786.10 cm$^{-1}$ above the  $a^3\Sigma^+$ $v=37$ state. In (c), the region between the $v=0$ and $v=1$ resonances is shown to have no additional features for direct detection. The dotted lines indicate the predicted dump laser frequency for the desired transition. Reprinted with permission from Sage {\it et al} \cite{sage2005}.}
\label{fig:sage}
\end{figure}

A first evidence of ultracold dimers obtained in the $v=0$ ground-state level has been reported for K$_{2}$ molecules in ref.~\cite{nikolov2000} (see Sec.~\ref{sssec:pa}). The formation of RbCs molecules in the $v=0$ ground-state level, while spread over several rotational states, has been clearly demonstrated in ref.\cite{sage2005}. Starting from dimers obtained through PA and spontaneous decay into high vibrational levels of the $a^{3} \Sigma^{+}$ lowest triplet state, the population of the $v'=37$ level has been transferred into the $v=$0 level of the $X ^{1}\Sigma^{+}$ ground state by an optical transfer process. Such a triplet-singlet conversion relies on the transfer through the $c ^{3}\Sigma^{+}$, $B ^{1}\Pi$ and $b ^{3}\Pi$ (correlated to Rb($5S$)+Cs($6P$)) molecular states coupled by spin-orbit interaction, allowing to circumvent the triplet-singlet electric-dipole-forbidden transition rule. Two different lasers generated the pump and dump pulses, while the molecules were detected by REMPI. The spectrum of the ion signal as a function of the dump laser frequency presented a peak corresponding to the position of the $v=0$ state, as shown in Fig.~\ref{fig:sage}. The estimated efficiency of the optical transfer from the $v'=37$ $a ^{3}\Sigma^{+}$ level into the $v=$0 $X^{1}\Sigma^{+}$ level was found around 6\%.

Formation of LiCs molecules in the rovibrational ground singlet state has been recently reported \cite{deiglmayr2008}. A single PA step through the $v'=4$, $J'=1,2$ levels of the $B ^{1}\Pi$ state (correlated to Li($2s$)+Cs($6P$)) populates the $X^{1}\Sigma^{+}$ $v=0$ level with a branching ratio of about 23\%. The detection is performed by REMPI through the $B ^{1}\Pi$ state. As this detection does not resolve the rotational structure, depletion spectroscopy with an additional narrow-band c.w. laser is implemented showing the presence of resonances corresponding to LiCs molecules in the absolute rovibronic level $v=0$, $J=0$ (after the decay of the $v'=4$, $J'=1$ PA level) and in the $v=0$, $J=2$ level (after the decay of the $v'=4$, $J'=2$ PA level). The estimated formation rates is 100~molecules per second and 5000~molecules per second in the $J=0$ and $J=2$ levels respectively.

An original approach based on optical pumping with shaped laser pulses has been demonstrated in \cite{viteau2008} to create Cs$_{2}$ molecules in the ground state $v=0$ level. The dimers are initially formed in low $X^{1}\Sigma^{+}_g$ vibrational levels ($v=1$ to 7) through PA of a manifold of spin-orbit-coupled $1_{g}$ states correlated to the $6S+6P$ and $6S+5D$ asymptotes, followed by two-photon radiative decay. Cs$_{2}$ molecules are detected by REMPI through the $C ^{1}\Pi_{u}(6s+5D)$ state. A broadband femtosecond laser pulse tuned to the transitions between the $X^{1}\Sigma^{+}_{g}$ and the $B ^{1}\Pi_{u}$ levels modifies the vibrational distribution of the ground state. If the laser pulses are shaped in order to cut the high frequency part of their spectrum and to avoid the repumping the $v=0$ molecules, the population finally accumulates into this level. The formation rate in the $v=0$ level is increased up to 10$^5$~s$^{-1}$, while the rotational distribution is not resolved. This method acts as a vibrational cooling process and  could be used in principle to cool the rotational degrees of freedom. It may be a general approach for any molecule providing that suitable electronic transitions can be found, with the drawback of heating the molecular sample. We also note that by appropriately shaping the pulses it is also possible to transfer the population to a selected vibrational level different from $v=0$ \cite{sofikitis2009}.

Another way to create molecules in the lowest rovibronic state combines the production of Feshbach molecules with the optical transfer technique known as STIRAP (Stimulated Raman Adiabatic Passage). A counter-intuitive sequence of laser pulses  fulfilling the adiabaticity criteria is set up, where the first pulse couples two molecular bound levels and the second one makes the coupling with the initial Feshbach molecular state. The process requires phase coherence between the two laser fields. The population flows through a dark state, avoiding radiative losses in the excited state. In these experiments the molecular fraction is detected by repeating the STIRAP process in the reverse sequence, and dissociating the Feshbach molecules. In this way it is possible to coherently couple the initial atomic sample to the molecular one and backwards without significantly heating the sample, thus allowing sensitive atomic imaging techniques to probe the molecular fraction. By making use of STIRAP, rubidium molecules produced through a Feshbach resonance in an optical lattice have been transferred to the second-to-last vibrational level (with a binding energy $\approx$500~MHz), with an efficiency of 87 \% \cite{winkler2007}. Using rf-transitions, the molecules have been transferred down to more deeply bound states \cite{lang2008}. The group at JILA has applied the same technique to $^{40}$K$^{87}$Rb heteronuclear Feshbach molecules, transferred to the third-to-last vibrational level (with a binding energy $\approx$10~GHz) with an efficiency larger than 80\%.

\begin{figure}[htb]
\begin{center}
\includegraphics[width=13cm,height=10cm]{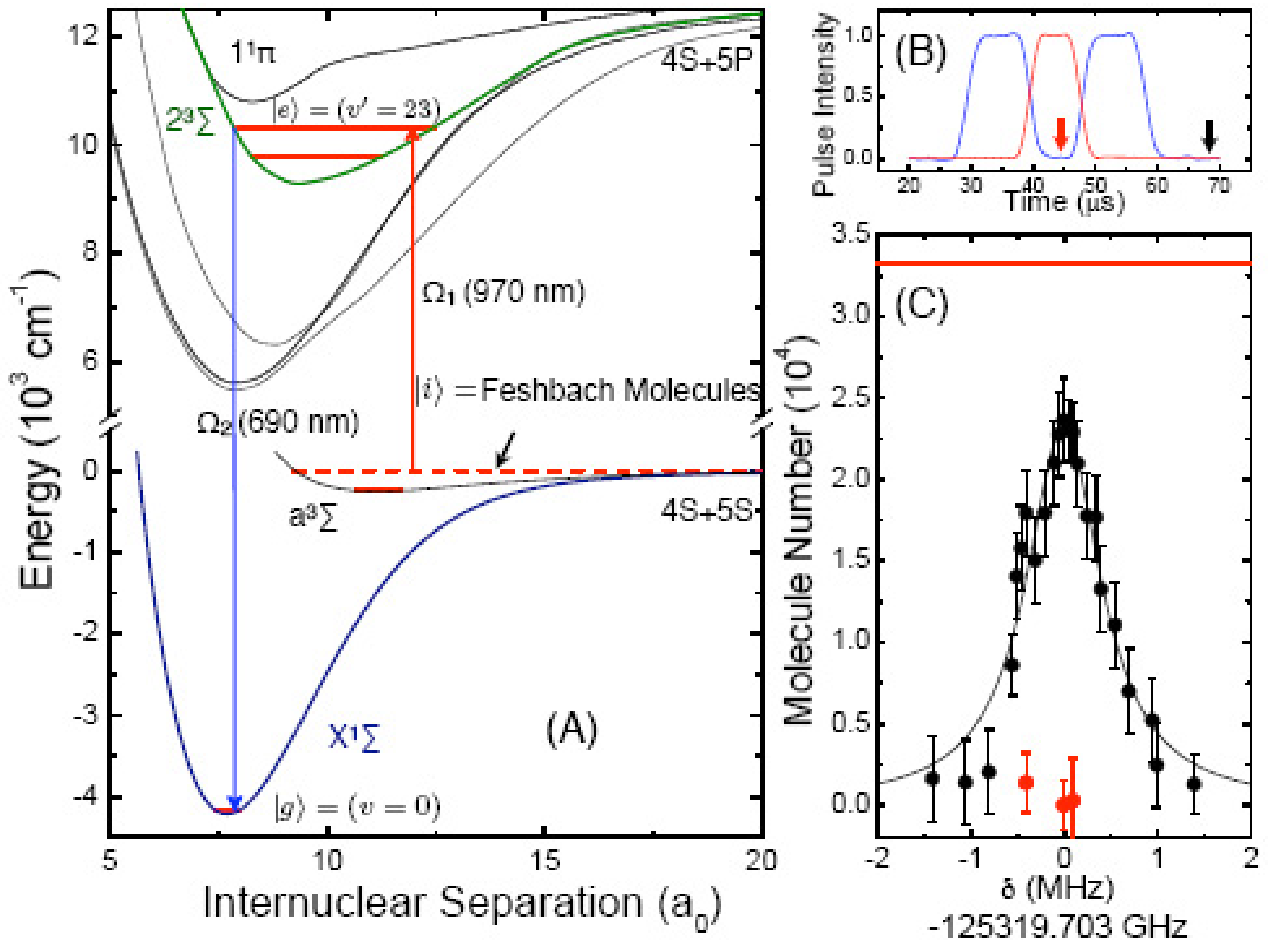}
\end{center}
\caption{STIRAP transfer from weakly-bound Feshbach molecules to the absolute molecular ground state (v = 0, N = 0 of X $^1$$\Sigma$). A: Transfer scheme, the intermediate state being the $v = 23$ level of the $\Omega$ = 1 component of the electronically excited (2)$^3\Sigma^+$ potential. B: Normalized Raman laser intensities vs. time for the round-trip STIRAP pulse sequence.  C: STIRAP line shape. The number of Feshbach molecules returned after a round-trip STIRAP transfer is plotted as a function of the two-photon Raman laser detuning. The round-trip data were taken at the time indicated by the black arrow in (B). The red data points
show the Feshbach molecule number when only one-way STIRAP is performed (at the time indicated by the red arrow in (B)), where all Feshbach molecules are transferred to the ground state and are dark to the imaging light. Reprinted with permission from Ni et al \cite{ni2008}.}
\label{fig:ni}
\end{figure}

Shortly afterwards, a major breakthrough has been demonstrated with the conversion of Cs$_2$ Feshbach molecules into a ground state level bound by more than 1000~cm$^{-1}$ \cite{danzl2008}. Lang {\it et al} succeeded transferring Rb$_2$ Feshbach molecules to the $a^3\Sigma_u^+$ lowest level \cite{lang2008a}, and similar results have been soon reported by the JILA group for both $X^1\Sigma^+$ and $a^3\Sigma^+$ KRb molecules \cite{ni2008} and in Innsbruck for $X^1\Sigma_g^+$ Cs$_2$ molecules \cite{danzl2008a}. In ref.\cite{lang2008a} the intermediate excited level of the STIRAP scheme belongs to the $(1)^{3}\Sigma_{g}^{+}(5S+5P)$ potential with $1_g$ character. The exact position of the resonance and the Rabi frequencies are obtained by accurate spectroscopy measurements that include atom-molecule dark resonances \cite{winkler2005}. The transfer has been done in a 3D optical lattice, that allows to obtain a lifetime for a fraction of the ground state molecules (those in the lowest Bloch band) exceeding 200~ms. In \cite{danzl2008a} the transfer down to the deeply-bound levels of the $X^1\Sigma_g^+$ state has been achieved by a double STIRAP transfer: the first step brings the population  into the $v=$73 level, and the second down to the $v=$0 level, using in both steps intermediate levels belonging to the $A ^{1}\Sigma_{u}^{+}$- $(b) ^{3}\Pi_{u}$($0_{u}^{+}$) coupled states. In \cite{ni2008} the $v=0$ levels of both the $a^3\Sigma^+$ and $X^1\Sigma^+$ states of polar $^{40}$K$^{87}$Rb have been reached with two different STIRAP paths. In the former case, the $v=$0 level was reached from Feshbach molecules through the $v'=$10 level of the $(2)^{3}\Sigma^{+}$ state with an efficiency of 56 \%. The lifetime of the produced molecules was measured at 170~$\mu$s, presumably limited by collisions with background atoms. In the latter case, the $v=$0 level was reached with the intermediate $v' =$23 level of the $\Omega =$1 component of the electronically excited $(2)^3\Sigma^+$ potential, coupled to the $B^{1}\Pi$ state through spin-orbit interaction, with a conversion efficiency of 83 \% (see Figure \ref{fig:ni}). The obtained polar molecules were then trapped in the optical dipole trap. The polar character of the dimers was confirmed by a measurement of the dipole moment through the Stark shift induced by a DC electric field on the dark resonance spectroscopy. The dipole moment has been measured to at 0.05~D for the $v=0, J=0$ level of the lowest triplet state and 0.566~D for the $v=0, J=0$ level of the singlet ground state. Following studies \cite{ospelkaus2009} measured also the AC polarizability of the ground state molecules at 1090~nm. The lifetime of the trap was dependent on the presence of K atoms in the dipole trap, as the reaction KRb+K$\rightarrow$K$_2$+Rb is energetically allowed. However, also removing the K atoms, the lifetime was limited to 70~ms for reasons still to be clarified. The ground state KRb molecules are actually just a factor 3 from quantum degeneracy.

\subsection{Decelerating and guiding molecules with external fields}

A method to directly cool molecules can use their interaction with external fields. It is already known since a few decades \cite{parker1989} that the inhomogeneous electric fields of multipolar devices can be used to manipulate (deflect, focus and orient) beams of dipolar molecules. Recently the interaction of molecules with electric, magnetic and optical fields has been used also to decelerate molecular beams. In the following subsections we describe the main techniques and results so far obtained. A specific review can be found in \cite{meerakker2008}.
	
\subsubsection{Stark deceleration of polar molecules}
	
\begin{figure}[htb]
\begin{center}
\includegraphics[width=10cm,height=8cm]{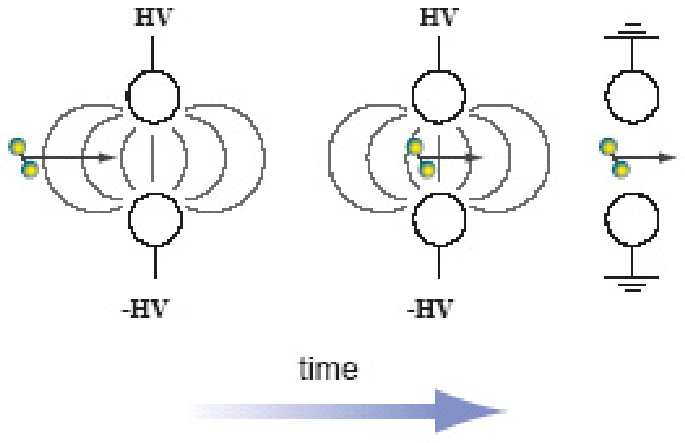}
\end{center}
\caption{A polar molecule having its dipole oriented antiparallel to the electric field lines will decelerate while flying into an electric field.  If the
electric field is switched off while the molecule is still in the field, the molecule will keep a lower velocity. Reprinted with permission from Bethlem et al \cite{bethlem2003}.}
\label{fig:stark0}
\end{figure}

A clever way to apply the Stark effect to decelerate polar molecules was demonstrated in \cite{bethlem1999}. The method does not produce real cooling because the phase-space density is preserved. It works with a supersonic molecular beam, where in the pulsed gas expansion efficient cooling of all internal degrees of freedom of the molecule occurs \cite{scoles1988}. Pulsed molecular beams have low rotational temperatures (few K) and densities of about 10$^{12}$ cm$^{-3}$ per quantum state. In the moving frame of the molecular beam the translational temperature is already quite cold (of the order of 1 K) and it is important for many applications, like for trapping, to transfer that to the laboratory frame. The deceleration process makes use of the interaction of dipolar molecules with time-varying electric fields. For many small molecules, positive Stark shifts of typically 1~cm$^{-1}$ can be obtained in an electric field of 100~kV cm$^{-1}$. Starting with molecule in a quantum state feeling an increase in Stark energy with increasing electric field (low-field seeker), the molecule will be decelerated while moving from low to high electric field. Leaving the high electric field region, the molecule will regain kinetic energy. This does not happen if the electric field is switched off quickly, as shown in Figure~\ref{fig:stark0}. In order to repeat the process many times, a Stark decelerator was built \cite{bethlem1999} consisting of many stages (64 pairs of electrodes). The Stark decelerator (that may work also as accelerator) is a device for neutral molecules equivalent to a linear accelerator for charged particles. The timing of the electric field switches defines a 'synchronous' molecule and determines the efficiency of deceleration and the stability of the process. Molecules slower than the synchronous molecule will see the field switched off before arriving to the position of the maximum, and will be less decelerated, while faster molecules are ahead and feel a higher deceleration \cite{bethlem2000a}. Therefore the process in the decelerator selects a bunch of molecules with a narrow velocity distribution, corresponding to a translational temperature of few mK, out of the starting distribution and decelerates it to an arbitrarily low absolute velocity. Moreover, the molecules in low-field seeking states are confined near the symmetry axis of the decelerator, where the electric field is minimum. In the first experiment \cite{bethlem1999} the metastable CO molecules were decelerated from 230~m.s$^{-1}$ to 98~m.s$^{-1}$. The molecules were detected by the electrons emitted impinging on a gold surface at the exit of the decelerator, with the time-of-flight distribution showing the deceleration process. Other molecules have been decelerated using Stark decelerators, including ND$_{3}$ \cite{bethlem2000}, OH \cite{bochinski2004,meerakker2005}, formaldehyde \cite{hudson2006}, NH \cite{meerakker2006} and SO$_2$ \cite{bucicov2008}. SO$_2$ molecules can be dissociated at threshold in a controlled way by external field \cite{jung2006}, and the fragments (O and SO) could be used for cold collision studies.

An envisioned application of Stark decelerated molecules is to yield an ensemble of slow molecules which could be trapped and further cooled down by sympathetic cooling with ultracold alkali atoms, thus extending the class of molecules which could be used for quantum degeneracy studies. Preliminary theoretical investigations performed on OH or NH molecules colliding with ultracold Rb atoms \cite{lara2006,lara2007,tacconi2007} are not favorable, as inelastic collisions are expected to dominate the process. The light LiH molecule has a quite simple structure, and could be an alternative for performing accurate modeling of collisional cross sections. A beam of Stark decelerated LiH molecules has just been produced using a 100 stages Stark decelerator, after transferring the population of the rotational ground level (which is a high-field-seeking state) into the first excited rotational level \cite{tokunaga2009}. Starting from a supersonic beam at 420 m/s velocity, the molecules have been decelerated down to 53 m/s, thus removing 98.5\% of the kinetic energy.

Another experiment demonstrated also deceleration of high-field seeking metastable CO molecules \cite{bethlem2002}. The scheme utilized alternating electric field gradients along the two directions perpendicular to the molecular beam in order to achieve guiding. Even molecules as heavy as benzonitrile (C$_7$H$_5$N) \cite{wohlfart2008} and YbF \cite{tarbutt2004} high-field seeking molecules have been decelerated. YbF molecules are of interest for a measurement of the electric dipole moment of the electron, that gains sensitivity due to the heavy mass of the molecule \cite{hudson2002}.

\begin{figure}[htb]
\begin{center}
\includegraphics[width=10cm,height=10cm]{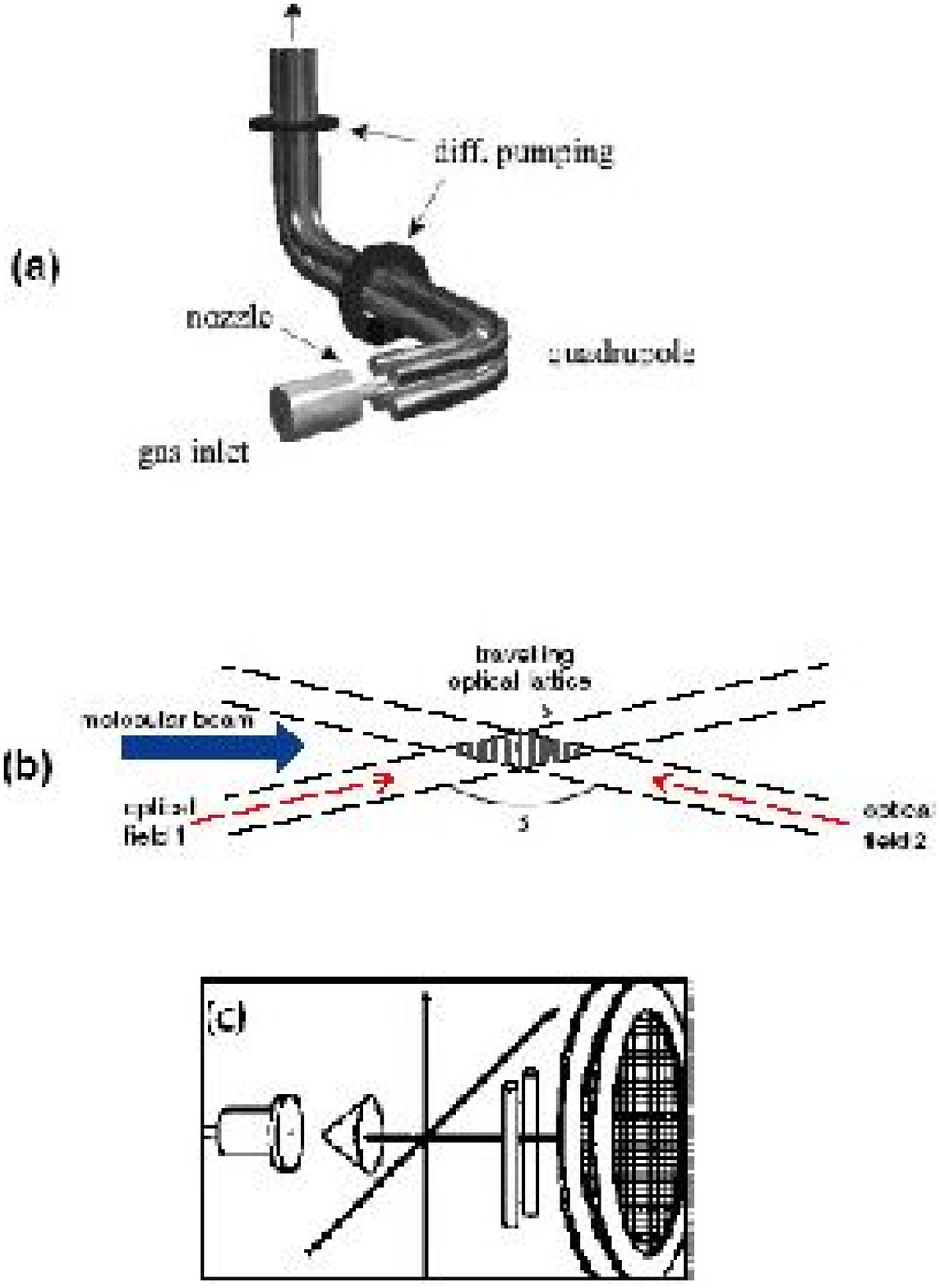}
\end{center}
\caption{Sketch of different methods using Stark effect to manipulate molecules. (a): velocity selection and guiding of polar molecules. Reprinted with permission from Junglen et al \cite{junglen2004}. (b): far off-resonant traveling optical lattice. Reprinted with permission from Fulton et al \cite{fulton2006a}. (c): scheme to decelerate H$_2$ molecules in Rydberg states. Reprinted with permission Yamakita et al \cite{yamakita2004}.}
\label{fig:stark1}
\end{figure}

An even simpler scheme to select and guide polar molecules using the linear Stark effect in an inhomogeneous field has been implemented in \cite{rangwala2003} and it is shown in Figure~\ref{fig:stark1}a. Using a bent guide with a quadrupole electric field, a beam of either formaldehyde (H$_2$CO) or deuterated ammonia (ND$_3$) has been guided and filtered in velocity, with a resulting longitudinal temperature of about 5 K and a transversal one of 0.5 K. Simultaneous guiding of molecules in high-field and low-field seeking states has been achieved \cite{junglen2004}. Using a four-wire setup	and switching the voltages of a pair of opposite electrodes, guiding of ND$_3$ molecules in both states was fulfilled for a range of driving frequencies.

A method to use the linear Stark effect to decelerate hydrogen molecules in Rydberg states has been demonstrated in \cite{yamakita2004}. The advantage of Rydberg molecules is their huge dipole moment compared to ground state molecules. There are however important disadvantages due to the finite lifetime (in the $\mu$s range for $n=$15-20) and to the complex Stark map in electric field and related state crossings. The experiment started from a supersonic beam of H$_2$, that were excited to selected Rydberg states by a combination of VUV and UV radiation. Two rods act as an electric dipole producing an inhomogeneous electric field along the axis that decelerates low-field seeking dimers and accelerates high-field seeking molecules, as sketched in Figure~\ref{fig:stark1}c. The observed deceleration corresponded to a decrease of kinetic energy of just 2.5 \% after the application of the dipole field for 1 $\mu$s. The use of many stages as for ground state molecules is in this case not applicable because the molecules would decay from the Rydberg state along the trajectory. A scheme using two dipoles was proposed in \cite{softley2005} to decelerate hydrogen molecules in Rydberg states to zero velocity, but it has not been implemented up to now.

\subsubsection{Zeeman deceleration of paramagnetic molecules}

Exploiting the Zeeman effect instead of the Stark effect, a magnetic analogue of the Stark decelerator has been realized. This allows to decelerate a wide range of molecules, possessing a magnetic dipole moment, to which the Stark technique cannot be applied. High magnetic fields with fast switching times have been difficult to achieve and required specific design of the coils. A first experimental demonstration of a Zeeman decelerator was firstly reported on hydrogen atoms in \cite{vanhaecke2007} using six stages, with a field distribution  providing a transverse restoring force to the beam axis. The application of this technique to molecules has been recently implemented \cite{narevicius2008}. Oxygen molecules from a supersonic beam were decelerated to velocities as low as 50~m.s$^{-1}$ using a decelerator with 64 stages. Oxygen molecules in the $^3$$\Sigma_{g}^{-}$ ground state are paramagnetic, but at high magnetic fields there are avoided crossings between different rotational states that can change character of the dimers from low-field to high-field seeking. As in the setup the high-field seekers were magnetically defocused, the magnetic field had an upper limit of operation. The velocity of $^{16}$O$_{2}$ molecules was directly measured using a quadrupole mass spectrometer mounted on a translation stage. An alternating gradient decelerator, like for the Stark method, can be used to decelerate diamagnetic species and high-field seeking molecules as well.

\subsubsection{Optical deceleration of polarizable molecules}

Deceleration using the second order Stark shift in intense far off-resonant optical fields has been demonstrated \cite{fulton2004}. A supersonic beam of benzene seeded in xenon interacted with an injection seeded Nd:YAG laser propagating in the orthogonal direction and focused to 20~$\mu$m, with a pulse width of 15~ns and a peak intensity above 10$^{12}$~W.cm$^{-2}$. The velocity was measured by ionizing the benzene molecules through REMPI by an UV laser and measuring the time-of-flight of the ions in a Wiley-McLaren mass spectrometer. The interaction with the radiation produced a deceleration or acceleration of the molecular beam following their position downstream or upstream of the pulsed laser respectively. The maximum deceleration corresponded to 25~m.s$^{-1}$. Pulsed optical lattices have also been used to decelerate molecules in a cold molecular beam \cite{fulton2006}. Nitric oxide molecules were decelerated by a deep optical lattice created by two near-counterpropagating laser beams with a fixed frequency difference, resulting in a lattice moving with constant velocity (Figure~\ref{fig:stark1}b). The molecules are temporally trapped within the lattice potential and make oscillations. If in the laser pulse time (6~ns) the molecules make an half oscillation in the potential well, corresponding to a half rotation in the phase space, their final velocity is determined by about twice the difference between the lattice and the molecular beam velocities. Therefore they can be decelerated (or accelerated) if the molecular beam velocity is greater (or smaller) than the lattice velocity. In the experiment NO molecules were decelerated from 400~m.s$^{-1}$ to 270~m.s$^{-1}$ in a single laser pulse \cite{fulton2006}. Molecules with a larger polarizability like benzene could be decelerated to zero velocity with appropriate pulse duration \cite{fulton2006a}.

\subsection{Kinematic cooling of molecules}

\subsubsection{Buffer gas cooling of molecules}

\begin{figure}[htb]
\begin{center}
\includegraphics[width=12cm,height=12cm]{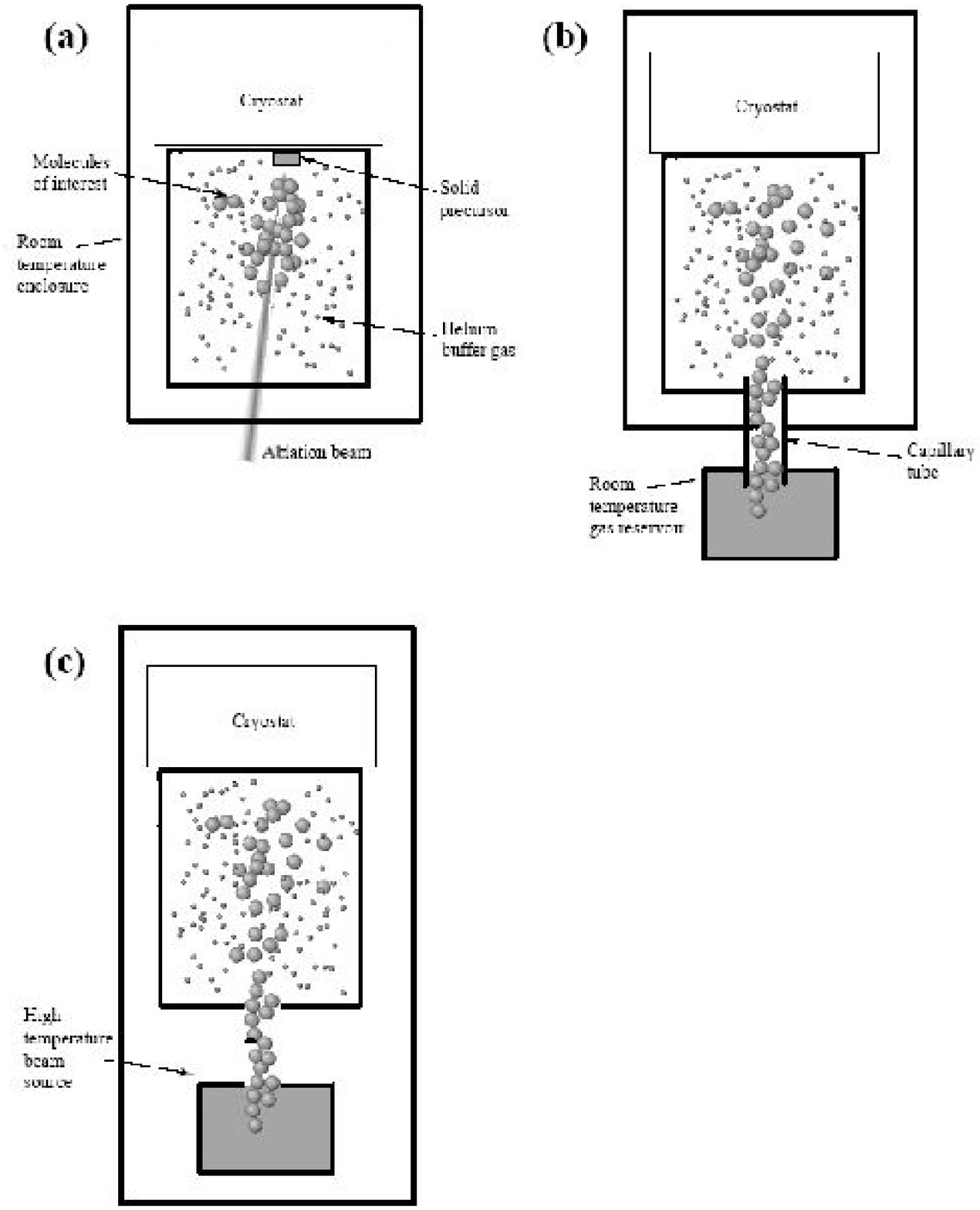}
\end{center}
\caption{Different methods to load a buffer gas cell for cooling and trapping molecules. Reprinted from Egorov \cite{egorov2004a}.}
\label{fig:egorov}
\end{figure}

The method of buffer gas cooling relies on the elastic collisions between the molecules (or atoms) and a cryogenic cooled helium gas. The method is quite general and in principle can be applied to any kind of molecule. It is a particular case of the sympathetic cooling method, that was firstly applied to atomic ions \cite{larson1986} and molecular ions \cite{baba1996}, and then to neutral atoms \cite{myatt1997}. In the first experiment with molecules \cite{weinstein1998a}, the molecular sample to be cooled was produced by laser ablation of a solid target with a starting temperature of about 1000 K, as shown in Figure~\ref{fig:egorov}(a). A buffer gas density of about 10$^{16}$ cm$^{-3}$ is available using $^4$He at a temperature of 800~mK or using $^3$He at 240~mK. With this density and by assuming a cross section for elastic scattering of 10$^{-14}$ cm$^2$, about one hundred collisions are necessary to thermalize close to the buffer gas temperature. Once the collisions with the buffer gas have dissipated the translational energy of the molecules, they can be trapped by a magnetic field. This step can be applied just for paramagnetic molecules in low-field seeking states, and in particular for those having an important magnetic dipole moment ($\mu\geq\mu_{B}$, where $\mu_{B}$ is the Bohr magneton). A spherical quadrupole field can be used to magnetically trap molecules, although it suffers for spin-flip (Majorana) losses in the center. In the first experiment \cite{weinstein1998a}, CaH molecules were produced by pulsed laser ablation of a CaH$_2$ solid target inside the cryogenic cell. The cell was positioned inside a magnet consisting of two superconducting solenoids arranged in anti-Helmoltz configuration, that could produce a maximum magnetic field of about 3 T, corresponding to a trap depth of 2~K for CaH molecules ($\mu$= 1 $\mu_{B}$). Vacuum separated the magnet (at 4~K) from the cell whose temperature (100-800~mK) was controlled by a dilution refrigerator and resistive heating. The cell was heated before the ablation pulse and successively cooled in order to pump the helium gas to the cell walls after the thermalization process. The loading process worked even without heating, presumably due to evaporation of the condensed helium gas by the ablation pulse. The trapped molecules were detected by laser induced fluorescence (LIF). The laser excited the $B^2\Sigma^+ v'=0 \leftarrow X ^2\Sigma^+ v''=0$ transition and the radiative decay $B^2\Sigma^+ v'=0 \rightarrow X ^2\Sigma v''=1$ was used for detection. In absence of the magnetic field a single rotational transition ($N^{\prime}=0, J^{\prime}=3/2\leftarrow N^{\prime\prime}=0, J^{\prime\prime}=1/2$) was observed, due to fast rotational relaxation. In presence of the magnetic field, the rotational transition splits into two shifted components, one coming from the low-field seeking state ($N^{\prime\prime}=0, J^{\prime\prime}=1/2, M^{\prime\prime}=1/2$), and the other from the high-field seeking state. By following the time evolution of the two components it was possible to observe the dynamics of the loading process and of the trapping. The low-field seeking state molecules had a trapping lifetime of 0.5~s. About 10$^8$ CaH molecules were trapped at a temperature of 400~mK. In a successive experiment about 10$^{12}$ PbO molecules were produced at a temperature of 4~K after pulsed laser ablation and thermalization in a cryogenic cell \cite{egorov2001}. PbO is one of the best candidates for measurement of a permanent electric dipole moment (EDM). CrH and MnH molecules have been cooled and trapped with the same technique, allowing the study of elastic and inelastic collisions with $^3$He atoms \cite{stoll2008}. A different loading mechanism using a molecular beam has also been developed \cite{egorov2004} (see Figure~\ref{fig:egorov}c). A radical beam source, consisting of a room temperature glow discharge in a mixture of gas, that converts ammonia molecules into NH radicals, produced a molecular beam entering the cryogenic cell through an orifice. Two shields, made by charcoal coated copper tube held at 4~K, pumped the helium leaking out of the orifice. The molecules loaded in the cell were detected by LIF and absorption spectroscopy on the $A^3\Pi_i (v'=0) \leftarrow X^3\Sigma^- (v''=0)$ transition. Up to 10$^{12}$ NH molecules were loaded inside the buffer gas cell at a temperature below 6~K. Recently, using the beam loading technique, 10$^8$ NH molecules were magnetically trapped with a $1/e$ lifetime of 200~ms for low-field seekers \cite{campbell2007}. NH molecules have also been trapped  together with atomic nitrogen in order to study cold chemical processes in conditions close to interstellar ones \cite{hummon2008}. Using laser ablation and buffer gas cooling, it is also possible to extract a cold molecular beam from a hole in the cell, as demonstrated for PbO \cite{maxwell2005}.

\subsubsection{Cooling by helium clusters}
In the free jet expansion of helium, cold nanodroplets, consisting of large clusters, can be formed at a temperature of about 0.4~K for $^4$He and 0.15~K for $^3$He. The droplets can be seeded by atomic or molecular species passing them through a pickup cell, the attached molecules being in the lowest vibrational state and at low rotational temperature. Such a device is the flying version of the buffer gas cooling approach above. Helium nanodroplet isolation (HENDI) spectroscopy is nowadays a well established technique for investigating molecules which are not necessarily accessible in the free space, or to measure the influence of the helium droplet on the molecular levels in order to know if the molecule sticks at the surface, or is solvated inside the droplet. Specific information can be found in the extensive reviews of refs.\cite{toennies2005,stienkemeier2006}. In relation with laser-cooling experiments, many studies have been carried out with alkali species. When two alkali atoms are picked up by the droplet, they tend to form a dimer at the droplet surface, dissipating their internal energy by evaporation of helium atoms. As the lowest triplet state of alkali dimers is about ten times less bound than their ground state, dimers are mostly formed in the former one \cite{higgins1988}, giving access to the spectroscopy of the triplet electronic states which cannot be easily accessed in the gas phase. The resolution of such a spectroscopy is usually low due to the interaction with the droplet which broadens the absorption or the emission lines. However, it can reveal most of the properties of previously unknown molecular states \cite{bruhl2001,bunermann2004,mudrich2004,ernst2006}. The low resolution of this spectroscopic approach aiming at measuring transition energies can be compensated by the high resolution on the energy spacing between vibrational levels: when they are coherently coupled during an excitation with a femtosecond laser pulse \cite{schulz2001,claas2006,mudrich2008}, the energy spacings appear as beating frequencies in the time-dependent light emitted by the molecules either trapped on the droplet, or flying close to it. Alkali timers in their higher spin metastable state - the lowest quartet state - can even be formed on the droplet \cite{nagl2008}. This represents a promising way to study the spectroscopy of such species in various mixtures, to be compared to the few available computations on alkali trimers (see for instance ref.\cite{cvitas2006}).

\subsubsection{Cooling molecules via ``Billiard-like'' collisions}

Another method that uses collisions to produce cold molecules has been developed and  applied first to NO molecules \cite{elioff2003}. The technique utilizes crossed supersonic molecular and atomic beams and it is based on the kinematic collapse of the molecular velocity distribution for scattering with a given recoil velocity vector in the center-of-mass (COM) frame. In an atom-molecule collision, when the final molecule velocity in the COM frame is equal in magnitude and opposite in direction to the COM velocity, the resulting velocity in the laboratory frame vanishes. If the COM and the recoil velocities scale in the same way with the starting molecules velocity, an effective cooling can be achieved. In the experiment \cite{elioff2003} NO molecules in the X $^2$$\Pi_{1/2}$ $j=1/2$ state collided inelastically with argon atoms in crossed beam geometry producing cold NO molecules in the X $^2$$\Pi_{1/2}$ $j^{\prime}=15/2$ state. For detection, the molecules were ionized by REMPI and the produced ions were imaged to a position sensitive detector. A fraction of about 10$^{-5}$ of the molecules in the seeded beam, corresponding to a density of 10$^8$ cm$^{-3}$, was cooled down to a translational temperature of 406~mK. In an improved experiment \cite{strecker2008}, the same group was able to lower the effective temperature of the extracted NO molecules down to about 35~mK, corresponding to a mean velocity of 4.5~m/s.

Another method has been recently implemented for metal-halogen molecules \cite{liu2007}. It uses reactive scattering of an alkali metal beam (potassium in that case) and a pulsed beam of HBr molecules. In the COM frame the final velocity of the KBr products gets slow because of the small ratio of the product masses $m_{H}/m_{KBr}$. In order to produce slow velocities also in the laboratory frame, the COM velocity is regulated near zero by sending the two beams counter-propagating and having a speed ratio equal to the inverse of their masses. The products are detected by surface ionization on the hot surface of a Re ribbon after time-of-flight. The outcoming distribution has a temperature of about 15 K with a 7 \% fraction of the beam at velocities below 14 m.s$^{-1}$, that could be trapped thanks to the large electric dipole moment of KBr.

\subsubsection{Rotating nozzle}

A conceptually simple method to slow molecules is the use of a rotating beam source. If the gas exits from a hole near the tip of the rotor, the velocity in the laboratory frame is given by the vector sum of the gas velocity in the rotor frame and the rotor velocity. The device can increase or decrease the velocity following the direction of rotation. Some effects play a role in the scheme including the centrifugal enhancement of the gas density inside the rotor and the swatting of the slow molecules by the rotor itself. A prototype was tested in \cite{gupta2001} for both supersonic and effusive beams of O$_2$, CH$_3$F and SF$_6$ molecules. The detection process measured the time-of-flight distributions by a fast ion gauge or a quadrupole mass spectrometer. The obtained temperatures were of the order of few K, limited by the collisions with the background gas that produced a velocity-dependent attenuation of the beam.

\subsection{Trapping of neutral cold molecules}

In order to perform sophisticated experiments on ultracold molecular samples, exceeding short timescales, much effort has been done for the storage of molecules in different kinds of traps. The possible methods for confining molecules include electrostatic, magnetic and optical traps or combinations of them. Each kind of trap works for specific molecules: the electrostatic trap for polar molecules, the magnetic trap for paramagnetic molecules, while the optical trap is more universal. Magnetic trapping of paramagnetic molecules was already discussed in connection with buffer gas cooling. Other kinds of traps, like the microwave trap \cite{demille2004}, have been proposed but not yet experimentally realized.

A powerful method for trapping employs the optical dipole force, confining particles towards or away from the region of maximum intensity, following the detuning sign. Useful traps are FORT's (Far Off-Resonant Traps) and in particular the quasi-electrostatic optical trap (QUEST)\cite{grimm2000}, working at large red detuning, i.e. at a frequency $\omega$ much lower with respect to the first atomic resonant frequency $\omega_{0}$. The main advantage of QUEST's is its capability to trap atoms in every quantum states and also molecules~\cite{takekoshi1998}. A further advantage relies on the low photon scattering rates that can determine very long storage times. In addition these conservative traps create ideal conditions to study collisions (see Section \ref{ssec:collisions}).

The trap depth of a quasi electrostatic trap is given by: $U\left( r\right) =\frac{2\pi}{c}\alpha I(r)$, where $I(r)$ is the laser intensity and $\alpha$ is the atomic (or molecular) dynamic polarizability which, for $\omega\ll\omega_{0}$, is almost equal to the static dipole polarizability ($\omega=0$). For example the rubidium static polarizability is equal to 329 a$_{0}^{3}$, so by tightly focusing a 100 W CO$_{2}$ laser, it is possible to achieve trap depths exceeding 1~mK. The stronger gradient, and consequently higher depth, is in the radial direction, while in the axial direction the gradient is lower, and depends on the Rayleigh length. For this reason, it is convenient to send the focused laser beam along the horizontal direction in order to easily compensate gravity on the vertical one.

Since its first experimental demonstration \cite{takekoshi1995}, QUEST's have been developed in many laboratories and used to reach BEC in both rubidium~\cite{barrett2001} and cesium~\cite{weber2003}, quantum degeneracy for fermionic lithium~\cite{ohara2002} and more recently molecular condensates, as already described. The first molecular trap using a QUEST has been done on Cs$_{2}$ dimers created by three-body recombination in a Cs MOT \cite{takekoshi1998}, and later extended to molecules produced by PA \cite{fioretti2004,staanum2006,zahzam2006}. Molecules created through magnetoassociation are routinely produced and trapped in QUEST's or FORT's. As Feshbach molecules are formed at very low temperatures (in the nK region), the trap depth can be shallow. This permits also the use of red-detuned optical traps where the relatively low laser intensity does not produce significant heating through off-resonant photon absorption. As already discussed in Sect.~\ref{sssec:lattices}, molecules can be confined in optical lattices, that in three-dimensions constitute an array of microtraps, that can preserve them from collisional inelastic processes.

\begin{figure}[htb]
\begin{center}
\includegraphics[width=10cm,height=10cm]{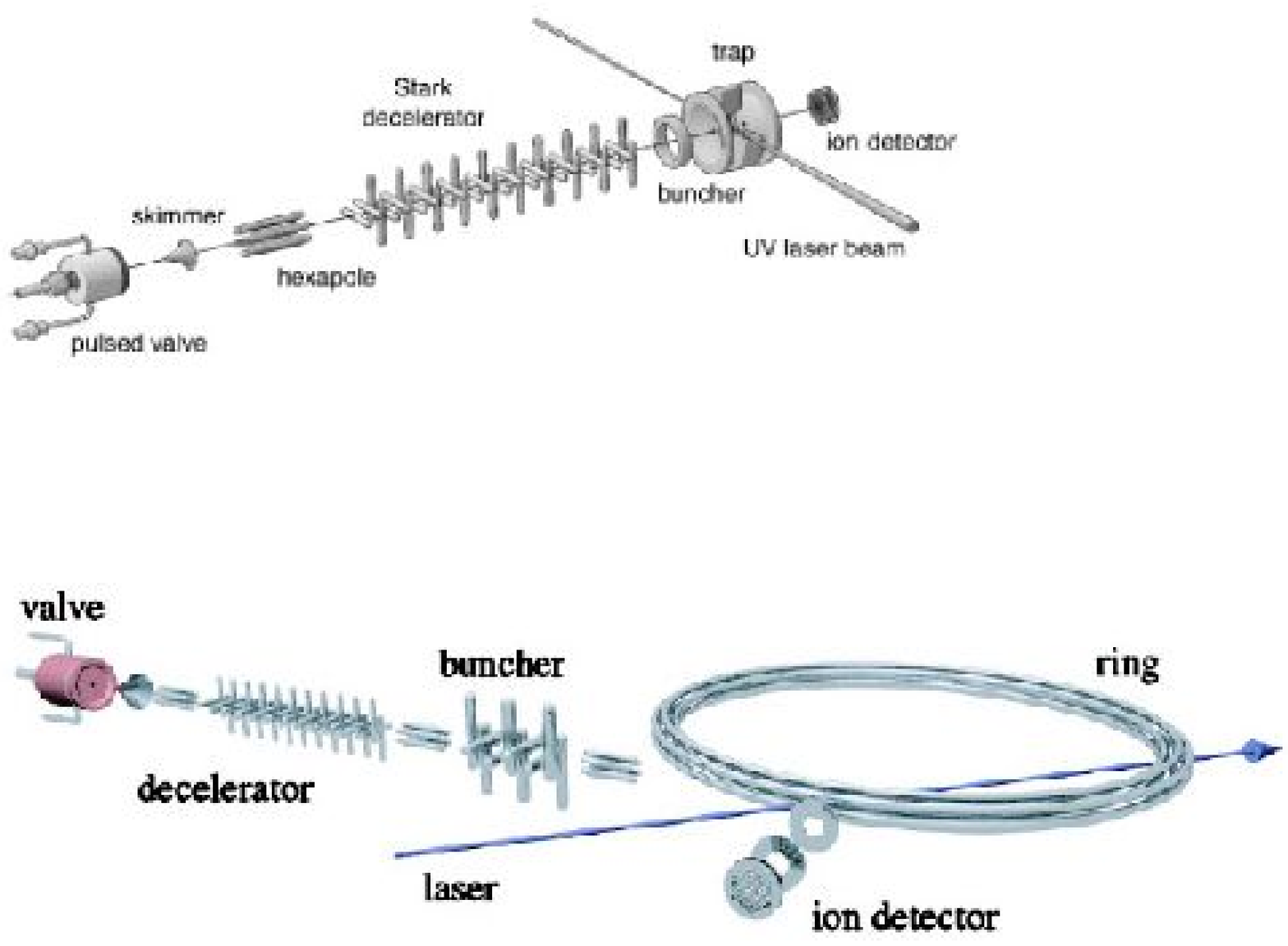}
\end{center}
\caption{Apparatus to trap Stark decelerated molecules in an electrostatic trap (above) and in a storage ring (below). Reprinted with permission from Bethlem et al \cite{bethlem2003} and from Crompvoets et al \cite{crompvoets2004}.}
\label{fig:stark}
\end{figure}

Polar molecules can be confined by suitable electric field configurations. As an extension of the Stark deceleration experiments, confinement of ND$_3$ molecules in an electrostatic trap has been demonstrated \cite{bethlem2000}. In the deceleration process molecules in the $J=1, K=1, M=1$ low-field seeking state of the vibrational ground levels were selected. The trap had a quadrupole geometry with a cut in the center along the decelerated beam axis (Figure~\ref{fig:stark}). The molecules entered the trap after being decelerated to 13~m.s$^{-1}$ with a velocity spread of 2~m.s$^{-1}$. By setting asymmetric voltages to the electrodes, the molecules were brought to standstill at the center of the trap. At that time, the voltages were switched to a near symmetric situation, giving a potential well in the trap center. The molecules were probed by multiphoton ionization and ion detection. The molecule density in the trap was 10$^6$ cm$^{-3}$ with a $1/e$ lifetime of 240~ms. In following experiments electrostatic trapping of OH \cite{meerakker2005} and NH \cite{hoekstra2007} has been achieved as well. When the trapping lifetime is improved, one must take in consideration the black-body radiation, as polar molecules have dipole allowed vibrational and/or rotational transitions in the IR region \cite{vanhaecke2007a}. Recently in a electrostatic trap of OH and OD molecules, the black-body pumping rates to high-field seeking rotational states have been measured, showing that room-temperature black-body radiation limits the trapping lifetime to a few seconds \cite{hoekstra2007a}. Electrostatic trapping of polar molecules has been achieved also under continuous loading by filtering the velocity of a beam of ND$_3$ molecules \cite{rieger2005}. In that case the molecular density was 10$^8$~cm$^{-3}$ with a lifetime of 130~ms.

Injection of ND$_3$ molecules in a storage ring with a radius of about 12 cm has been demonstrated \cite{crompvoets2001}. In this case the decelerated molecular beam with a velocity of 89~m.s$^{-1}$ was tangentially injected into the storage ring (see Figure~\ref{fig:stark}), passing between the rods of an hexapole, which was abruptly switched on once the molecules entered it. The inhomogeneous electric field provided the necessary centripetal force to produce stable orbits. In the ring bunches of cold molecules interact repeatedly, at well defined times and at distinct locations, with electric fields. Up to six round-trips inside the ring have been observed. A modified design of the storage ring, that introduces a gap between two half-rings, allowed to counteract the spreading of the molecular packet enabling a larger number of trips \cite{heiner2007}. In principle, different molecules can be simultaneously stored in the ring and used for collision studies.

Trapping of OH molecules in a magneto-electrostatic trap has been achieved in \cite{sawyer2007}. The OH molecules were decelerated in a 142 stages Stark decelerator before they enter the trap. The molecules are in the $\left|\Omega=3/2\right\rangle$ ground state. When high external fields ($E\geq$1~kV/cm and $B\geq100$~G) are applied, the total angular momentum \textbf{J} and its projection $m_{J}$ on the axis of the applied field become good quantum numbers; the ground state also splits into two opposite parity states ($\Lambda$-doublet). In these conditions the ground state molecules feel both linear Zeeman and Stark effects. The magneto-electrostatic trap is composed by a quadrupole magnetic field superposed to a quadrupole electric field. The molecules that are decelerated must be of electrically low-field seeking nature ($\left|J=3/2,m_{J}=\pm3/2\right\rangle$) while those which are aimed for trapping must also be of magnetically low-field seeking nature, therefore just the half in the $\left|J=3/2,m_{J}=+3/2\right\rangle$ state. The experiment succeeded to trap OH at a density of 10$^3$~cm$^{-3}$ and a temperature of 30~mK \cite{sawyer2007}.

By realizing an electrodynamic trap, the molecules in high-field seeking states have been trapped \cite{veldhoven2005}. Molecules in the absolute ground state are high-field seekers and it is important to be able to manipulate them. As they are not sensitive to inelastic collisions evaporative cooling can in principle be applied to increase their phase-space density. The Earnshaw's theorem forbids a stable extremum of the electromagnetic field in three dimensions. However it is possible to have a maximum in two dimensions and a minimum in the other. Switching between two configurations of electric fields with a saddle point can confine either low-field or high-field seeking molecules. The AC trap has been realized by combining a dipole field and an hexapole field and tested with ND$_3$ molecules in the $J=1, K=1$ ground state. After Stark deceleration of the low-field seeking hyperfine level, the molecules can be partly pumped to the high-field seeking state by a microwave pulse. The trap is stable for a switching frequency above a cut-off value (about 1~KHz), until for too high switching frequency the net force averages out. Details on the shape of the trap and motion inside it can be found in \cite{bethlem2006}.

Electric traps can also be used for confinement of polar molecules produced by photoassociation. The first demonstration has been given for NaCs dimers \cite{kleinert2007,kleinert2007a}, using a thin-wire electrostatic trap (TWIST). The molecules are produced by PA in a large number of deeply bound vibrational states ($v=0-30$) of the $X ^1\Sigma^+$ ground state. In presence of an electric field, they feel a quadratic Stark effect with a polarizability depending on the ratio between the electric dipole moment and the rotational constant. As both these two quantities decrease in a similar way as a function of the vibrational number, the polarizability is constant within 10 \% for the first thirty vibrational states. The dimers are high-field seekers for $J=0$ and low-field seekers for rotating states. The molecular trap is directly loaded by PA in a MOT using tungsten wires as electrodes in a quadrupole configuration. The trap confines about 100 molecules with a lifetime of 225~ms, limited by the background vacuum.


\section{The challenge for theorists: the knowledge and the control of cold molecule interactions}

One of the dreams of scientists is to control the evolution of a complicated quantum system at the ultimate limit, i.e. to control both its internal and external degrees of freedom. This implies mastering its formation process, and its interaction with similar neighboring systems, or with its environment. As it can be understood from the previous sections, cold molecules represent attractive systems to achieve such a goal, as many experimental groups were successful to create molecules in well-defined internal states and to trap them in elaborate electromagnetic configurations. Such progress clearly benefited from a continuous exchange between theory and experiment, either for predictions or interpretations of observed results.

\subsection{The knowledge of the structure of cold atom pairs: a cornerstone}

Researches on cold molecules mainly started with the formation of alkali dimers, as alkali atoms are the most widely used species for laser cooling. A detailed knowledge of their structure is necessary for appropriate manipulation of these species in the course of their formation process (association and stabilization) as well as their detection process (REMPI). The simple electronic structure of alkali atoms, i.e. a single valence electron in the field of a polarizable ionic core, allows for calculating the electronic properties of alkali dimers with the hope of reaching a somewhat high accuracy. Several popular quantum chemistry codes known by their acronyms like CIPSI \cite{huron1973}, MOLPRO \cite{knowles1989,werner2008}, GAUSSIAN \cite{frisch2003}, can be efficiently used for such systems. But it is virtually impossible to determine the absolute energy of molecular levels with an accuracy comparable to the extreme precision of the measurements yielded by molecular spectroscopy or cold molecule experiments.

A noticeable exception is the simplest neutral molecule, i.e. H$_2$ which is a pure two-electron system. Elaborated approaches based on explicitly correlated basis functions in elliptic coordinates have been specifically designed to take in account the deviations from the usual Born-Oppenheimer (BO) approximation, which are important in H$_2$ due to its relatively low mass \cite{wolniewicz1998}. Joint experimental and theoretical studies on several electronic states have demonstrated an accuracy better than 1~cm$^{-1}$ on the hydrogen molecular levels \cite{reinhold1999,lange2001}. The alternative approach of Multichannel Quantum Defect Theory (MQDT) has been also used for exploring the electronic and vibronic structure \cite{ross1994} of H$_2$ excited states, based on previous accurate {\it ab-initio} results like the ones mentioned above. Recently, spectacular results have been obtained for the hyperfine structure of the H$_2$ Rydberg states and its parent ion H$_2^+$, reproducing the position of hyperfine experimental energy levels to an accuracy better than 1~MHz \cite{osterwalder2004,cruse2008}.

Going back to the alkali dimers, such accurate studies are out of reach due to the intrinsic multielectronic nature of these molecules. However the relative simplicity of the alkali atoms, i.e. composed of one valence electron moving in the field of a polarizable core, stimulated many elaborate theoretical studies on alkali diatomic molecules. A large amount of high-resolution molecular spectroscopy studies have been carried out as well, and for both cases, their complete list would require an entire review article, or even a book. In the following, we attempt to provide the reader with a documented overview of their current knowledge, relevant for cold molecule studies. Most of the quoted references contain extensive bibliography which could help the interested reader to locate information on a specific system, or on specific molecular states.

As alkali species are involved in association processes from ultracold atom pairs, it is first appropriate to divide the range of interatomic distances in two regions. First in the long-range domain (say, $R >$20$a_0$ typically, where the exchange of electrons between atoms vanishes), the dynamics of the association of the atom pair is dominated by atom-atom interactions induced by electrostatic forces. As it is well-known \cite{landau1967,hirschfelder1967}, two ground-state $S$ atoms interact via second-order dispersion forces, leading to a potential energy varying as $-C_6/R^{6}$ where $C_6$ (of positive sign) is referred to as the van der Waals coefficient. Two identical atoms, one  being in its $S$ ground state, and the other in the first excited $P$ state feel each other via a first-order dipole-induced dipole interaction yielding a long-range potential energy proportional to $-C_3/R^{3}$. The $C_3$ coefficient may be positive or negative, depending on the relative orientation of the dipoles induced on each atom. Similarly, a ground state $S$ atom interacting with another identical atom in its first excited $D$ state gives rise to a first-order potential energy varying as $-C_5/R^{5}$ due to quadrupole-quadrupole interaction. Note however that in the latter case, this term is dominated by the Van der Waals interaction for most of the range of interest, due to the usually large values of the $C_6$ coefficients. Finally, two atoms of different species influence each other via van der Waals interaction, as there is no possibility for exchanging their excitation, i.e. for inducing permanent dipole or quadrupole dipole moments. In all the above situations, the potential energy of the atom pair is entirely determined by atomic properties. Due to their simple structure, highly accurate wave functions for the atoms can be derived, resulting into precise values of the $C_n$ long-range coefficients. Among many theoretical studies, we can quote the most precise ones concerning a pair of ground state atoms (including Li to Fr), yielding either $C_6$ values, and higher order terms ($C_8$, $C_{10}$ terms) \cite{yan1997,derevianko1999,porsev2003} for homonuclear pairs, and for heteronuclear pairs \cite{derevianko2001}. For a pair of one ground state atom and an excited atom, extensive studies can be found in refs \cite{bussery1987,marinescu1995,marinescu1999}. Let us note that similar studies have been performed for other species relevant for ultracold atoms like alkaline-earth species \cite{porsev2002,mitroy2008}.

Modern quantum chemistry computations accurately deal with electron correlations, which control the quality of the electronic wave functions. In this respect, several electronic properties which are often hardly measured accurately, can be determined from theory with a great confidence. A systematic investigation on all alkali pairs in the framework of Effective Core Potentials (ECP) and full Configuration Interaction (CI) \cite{aymar2005} has recently predicted the value of their permanent electric dipole moment in the ground state, which has been for instance recently confirmed in the course of an experiment creating ultracold KRb molecules in their $v=0$ level \cite{ni2008}. A discrepancy of about 40\% remains between three different theoretical models. Transition dipole moments, which drive the transition probabilities, can be accurately determined as well, and it is now proved that elaborate computations of different kinds yield satisfactory agreement among themselves \cite{aymar2007}. Similarly, the computation of static dipole polarizabilities
\cite{deiglmayr2008}, and dynamic polarizabilities \cite{kotochigova2006} are relevant for modeling the trapping of polar molecules in external electromagnetic fields, or to predict their possible orientation and alignment in a superposition of static and oscillating electric fields \cite{friedrich1999a}.

In the perspective of obtaining molecular quantum degenerate gases, ultracold molecules have to be created in their absolute ground state, or at least in a single quantum state. This implies the control of the hyperfine interaction in the low-lying molecular levels which are ultimately used in STIRAP-based transfer processes (see section \ref{sssec:v=0}). Aldegunde {\it et al} investigated the hyperfine energy levels and Zeeman splitting for homonuclear and heteronuclear alkali dimers in low-lying rotational and vibrational states, by carrying out density-functional theory calculations of the nuclear hyperfine coupling constants \cite{aldegunde2008,aldegunde2009}. For instance, rotationless levels are split into multiplets, where two neighboring levels are separated typically by amounts between 90~Hz for $^{41}$K$_2$ and 160~kHz for Cs$_2$.

While alkali atoms still represent the preferred species for ultracold molecule experiments, other species like alkaline-earth atoms \cite{nagel2005,yasuda2006,vogt2007}, or Ytterbium \cite{takasu2004,tojo2006,kitagawa2008}, are also used by several groups to perform PA spectroscopy, even if no stable ultracold molecules have been observed yet. Such atoms possess two valence electrons, so that the calculation of the electronic structure of the associated dimers will be more involved. In particular,  the corresponding dimers can be modeled as four-effective-electron systems within an ECP approach, which cannot be treated anymore using a full-CI computation, as the configuration space is by far too large compared to the capabilities of current or upcoming computers. Elaborate calculations are nevertheless feasible, as attested by the following studies achieved with the objective of finding efficient cold molecule formation schemes. The electronic structure of the calcium dimer, including potential curves, transition dipole moments and spin-orbit couplings, as well as the predicted PA spectrum, have been investigated in detail in refs.\cite{bussery-honvault2003,bussery-honvault2005}, using a hybrid method involving small-core ECP's. The electronic structure of the Sr$_2$ molecule, including potential curves for the ground state and for the lowest {\it ungerade} states, and transition dipole moments, has been obtained in an {\it ab initio} relativistic configuration interaction valence bond self-consistent-field method. The accuracy of the results have been checked against the size of the employed basis sets, and compared with available experimental data. For both systems, the possibility to stabilize photoassociated molecules into ground state molecules due to non-adiabatic couplings is discussed. Despite its two valence electron, the treatment of the Yb$_2$ molecule is even harder due to the large Yb$^{2+}$ polarizable core whose size is comparable to the size of the valence orbitals. Ground and excited state potentials have been computed with the MOLPRO package, using three different ECP's \cite{wang1998c}. It is worthwhile to note also that in the near future, data concerning mixed species will be required, as ongoing experiments are setting up cold atom traps with mixed species like for instance Yb and Rb \cite{nemitz2009}.

It is also worthwhile to mention a possible novel interest for alkali hydrides in the cold molecule context (see the recent Stark deceleration of LiH \cite{tokunaga2009}), whose electronic properties are similar to the one of alkali dimers, and which can be accurately calculated (see for instance refs. \cite{boutalib1992,geum2001,khelifi2002,khelifi2002a,zrafi2006,aymar2009}). Based on such knowledge, the possibility to create cold LiH and NaH ground state molecules has been investigated \cite{juarros2006,juarros2006a} by one-photon stimulated radiative association process applied on a pair of colliding cold Li and H atoms. Significant population of high-lying vibrational states is predicted, while the spontaneous decay cascade down to the $v=0$ level may take several minutes. Of course, such a proposal relies on the availability of a stable dual trap of cold Lithium and Hydrogen atoms.

Cold molecule studies benefited a lot from the high-resolution spectroscopy of alkali dimers. It is impossible here to list the abundant literature available on this class of systems which is intensively studied since almost 40 years. Such systems are indeed quite popular in molecular spectroscopy, for the same reason as in laser-cooling studies, i.e. most of the main transition frequencies lie in the optical domain. It is striking to realize that the cold molecule experiments performed with all possible pairs of alkali atoms boosted many new high-resolution spectroscopic investigations, which in turn were helpful to guide them. This is particularly true for heteronuclear alkali pairs as the ground state and the lowest triplet state of most of them have been recently determined in an amazing series of studies \cite{docenko2004,docenko2005,docenko2006a,staanum2007,ferber2008}. Moreover, several excited electronic states have been identified, including those exhibiting strong perturbations like the well-known case of the lowest excited $A^1\Sigma^+$ and $b^3\Pi$ states (also referred to as the ($A \approx b$) system) coupled by spin-orbit interaction. This effect is dominant when at least one heavy alkali atom (Rb or Cs) is involved \cite{amiot1999}. Such studies rely on elaborate deperturbation methods \cite{lisdat2001} which most often imply a joint theoretical and experimental analysis: initial guesses for potential curves and molecular spin-orbit coupling are provided by quantum chemistry calculations as input for a multi-parameter non-linear fitting procedure \cite{tamanis2002,bergeman2003,docenko2007,salami2009}. The input of the quantum chemistry calculations is even more important for instance in the case of the most recent experiments devoted to the production of $v=0$ molecules with STIRAP (see Section \ref{sssec:v=0}), for which the complete spectroscopy of the intermediate excited states was not available. Such experiments have a quite long duty cycle, so that the appropriate transitions cannot be identified by carefully looking for all of them. Therefore the calculations are helpful for the assignment and the interpretation of the few observed lines.

\subsection{Control of the formation of cold molecules with external fields}

In cold and ultracold gaseous samples, the kinetic energy of the particles is smaller than the perturbations induced by the presence of external electric and magnetic fields, which can therefore modify their dynamics. This offers the opportunity to achieve a long-lasting goal of controlling elementary chemical reactions, for instance to drive products into a single, well-defined quantum state. In this respect, ultracold conditions have the advantage of drastically reducing the number of channels involved in the dynamics, so that precise comparison between experiment and theoretical models can be envisioned.

\begin{figure}[htb]
\begin{center}
\includegraphics[width=10cm,height=10cm]{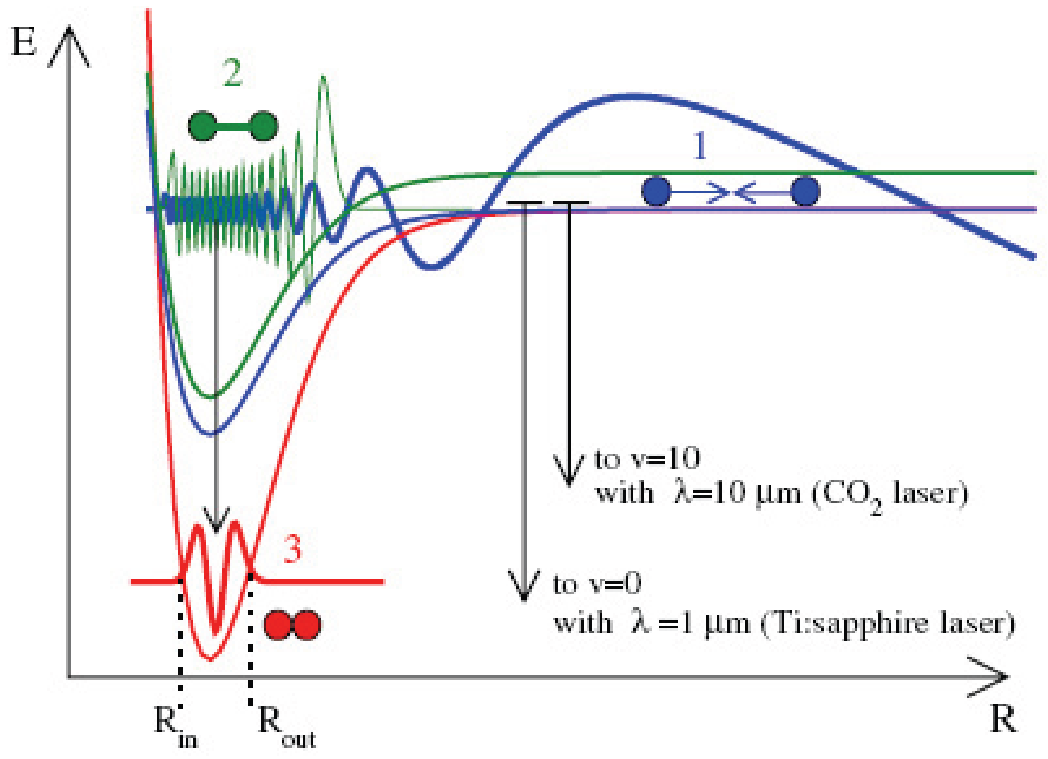}
\end{center}
\caption{The Feshbach-optimized photoassociation process (FOPA): Colliding atoms (1) interact via open (blue) and closed (green) channels due to hyperfine interactions. A Feshbach resonance occurs when a bound level
(2) (green wave function) coincides with the continuum state (blue wave function). A photon can associate the atoms into a bound level v (3) of the ground state potential (red). Reprinted with permission Pellegrini {\it et al} \cite{pellegrini2008}.}
\label{fig:fopa}
\end{figure}

As already described in section \ref{sssec:magnetoass}, external field assisted cold collisions are the basis for the magnetoassociation of ultracold atoms through the control of magnetic Feshbach resonances (MFR) by tuning an external magnetic field. Mastering the position of these resonances in an experiment requires a detailed knowledge of the related atom-atom interaction mainly at large distances, characterized by the Van der Waals coefficient and the scattering length, as previously discussed. A recent proposal suggested to use the enhancement of the density probability of the atom pair at short distances induced by the MFR to perform a radiative association to stabilize the MFR into the $v=0$ level of the molecular ground state \cite{pellegrini2008}. This process, named as Feshbach-Optmized photoassociation (FOPA) by the authors (Figure \ref{fig:fopa}), is predicted to provide molecular formation rate as high as a few $10^6$ molecules/s in the case of LiNa, comparable to what is observed in the experiments reported in section \ref{sssec:pa}.

Such a control of resonance positions can be extended to the optical Feshbach resonances (OFR), induced by a strong continuous resonant laser light, which couples open and closed scattering channels. The case of non-resonant laser light applied on cold atoms has been treated in ref. \cite{melezhik2003}. First predicted in ref.\cite{fedichev1996}, OFR's have been observed for the first time by Fatemi {\it et al} in 2000 \cite{fatemi2000}, in the photoassociation of cold Na pairs. OFR's have been proposed as a versatile tool to create cold molecules: indeed, they involve excited electronic states of molecules, instead of requiring the presence of an hyperfine manifold as for MFR's, which is for instance absent in most abundant isotopes of alkaline-earth atoms. Their tuning is also more flexible through both the laser intensity and frequency. The coherent conversion of an atomic Bose-Einstein condensate into a molecular one using photoassociation laser with a linear frequency sweep has been modeled \cite{drummond1998,javanainen1999a}, suggesting the possibility to observe oscillations in the atom/molecule number with time. Koch {\it et al} \cite{koch2005} modeled the formation of ultracold $^{87}$Rb$_2$ molecules by solving the time-dependent Schr\"odinger equation, involving a linear ramping of the laser frequency: in the dressed molecular state picture, the OFR is swept through the entrance scattering channel, so that ground state molecules are left in their highest vibrational levels.

For molecules exhibiting a permanent dipole moment, such ase heteronuclear alkali dimers, external electric fields also offer the possibility to tune the interaction within the atom pair \cite{marinescu1998b}. Strong electric fields have been shown to strongly distort the rovibrational internal structure of heteronuclear molecules. In refs.\cite{gonzalez-ferez2004,gonzalez-ferez2005}, the authors provided a full rovibrational description of the molecule in a homogeneous electric field including the coupling between the vibrational and rotational motions, through the development of an effective rotor approximation going beyond the traditional rigid rotor approximation. The same group applied this approach to the one-photon induced photoassociation of the strongly dipolar LiCs molecule \footnote{The LiCs molecule possesses the largest permanent dipole moment of all alkali pairs, $5.3$~Debye \cite{aymar2005}, in the $v=0$ of their ground state.}, under the influence of a static electric field in the range of $10^{-7}$ to $10^{-4}$~a.u. (or 0.514-514~kV/cm) \cite{gonzalez-ferez2007}. The cross-section has been found to strongly depend on the rotational angular momentum, with an inhibitory effect predicted on $J=1$ levels when the field increases.
Finally, the combination of an electric and a magnetic field on a heteronuclear dimer (LiCs) has been shown to induce Feshbach resonances for moderate electric fields (about 100~kV/cm), also inducing strong anistropy in the atom-atom interaction \cite{krems2006,li2007}.

Besides these studies, a wealth of theoretical developments have been carried out in order to apply the techniques of coherent control or optimal control with laser pulses to the formation of ultracold molecules, with the objective of controlling the association process with appropriate shaping of the pulse in amplitude and phase. The first attempt to model wave-packet dynamics of the PA of cold atoms has been proposed in ref.\cite{machholm1994}. A pair of cold sodium atoms was associated by a picosecond laser pulse, and then probed by a second pulse ionizing the associated pair. The objective was to discriminate the two possible ionization mechanisms, i.e. photoionization and autoionization, expected to yield different time dependent molecular ion count. The corresponding experiment was found to actually probe the dynamics of the wave-packet in the intermediate, photoassociated state \cite{fatemi2001}. Only very few further PA experiments with femtosecond laser pulses have been reported afterwards, probably because up to now they have not revealed a clear manifestation of the possibility to enhance and control the formation of cold molecules \cite{brown2006,salzmann2006}. On the theoretical side, the first attempts consisted in optimizing the photoassociation step using frequency-chirped laser pulses \cite{vala2001,luc-koenig2004c,marquetand2007}. A chirped broad pulse allows maximizing the number of excited pairs of atoms located at different interatomic distances, as resonant conditions for PA are then reached for many frequencies embedded in the pulse envelope ("photoassociation window"). A large variety of laser pulse has been examined \cite{luc-koenig2004c} for picosecond duration (which is adapted to the typical dynamical time of cold atom PA), and various detuning of their central width. The wave-packet describing the excitation of the cold atom pair in the photoassociated state is predicted to focus at the inner turning point of the excited molecular state. This yielded favorable conditions for stabilization by spontaneous decay into stable ground state molecules \cite{koch2006a,schafer-bung2006}, which has been modeled  through pump-dump schemes following the ideas of optimal control theory \cite{koch2004a,koch2006,poschinger2006}. Such studies are still at their beginning, but look very promising. However they are currently limited due to the possibility to include more channels in the calculations (to mimic the hyperfine structure for instance), and to properly describe the repetition of the pulse \cite{shapiro2007}, which could involve couplings with several partial waves in the entrance scattering channel. In this respect, modeling such processes on alkaline-earth species with no hyperfine structure \cite{koch2009} may be of interest for future experiments.

\subsection{Interactions between cold atoms and molecules}
\label{ssec:collisions}

Several recent experiments made giant strides in forming dense samples of ultracold molecules, either mixed or not with ultracold atoms, and started to address the issue of understanding their mutual interactions. First, much less channels are expected to contribute at ultra-low collision energies than in the thermal domain, as the rotational angular momentum $\ell$ of the complex is zero at vanishing collision velocities $v_{coll}$. The possibility to compare elaborate theoretical models to detailed experimental results obtained under well-controlled conditions is now at reach. In particular, the cross sections are governed by the Wigner's threshold laws \cite{wigner1948}, predicting elastic cross sections varying as $v_{coll}^{4\ell}$, and inelastic cross sections as $v_{coll}^{2\ell-1}$. In consequence, ultra-low energy collisions are dominated by the $s$-wave ($\ell=0$), and reaction rates can be quite large at ultracold temperatures. Next, the large de Broglie wavelength of the collision partners will enhance the importance of tunneling effects during the process. The long-range interactions between the collision partners will play a crucial role, as well as the coherences if the collisions occur within a degenerate gas. Finally, the collision process can be modified, and hopefully controlled, by the presence of external electric or magnetic fields, as the corresponding coupling terms are of the same order of magnitude, or even larger, than the collision energy. Such studies opened a new era for the vast field of inelastic or reactive collisions, which could be renamed as cold or ultracold chemistry \cite{krems2005,weck2006,hutson2006}.

In this section, we first review the experiments which reported evidence for atom-molecule or molecule-molecule collision phenomena at low temperatures, and we will overview the status of the theoretical developments.

\subsubsection{Observation of cold atom-molecule and molecule-molecule collisions}
\label{sssec:collisions_exp}

Two different experiments on optical trapping of Cs$_{2}$ dimers created by PA allowed to extract atom-molecule and molecule-molecule inelastic rate constants for different rovibrational states \cite{staanum2006,zahzam2006}. The rates turned out largely independent on the rovibrational level populated with corresponding cross sections unitary limited. Recently also inelastic collisions in a trapped sample of RbCs molecules have been studied \cite{hudson2008}. Atom-molecule collisions were studied for both RbCs-Cs and RbCs-Rb cases, by removing the undesired atom from the optically trapped sample. The REMPI state-selective detection allowed to measure rate constants for a range of vibrational levels of the ground triplet states with binding energies from 0.5 to 7~cm$^{-1}$. The inelastic rate constants were equal within the experimental precision. The rate behavior and magnitude can be explained by a model that assumes that any pair that penetrates the short-range region gives trap loss. The flux transmitted at short range is calculated by solving the Schr\"{o}dinger equation with the appropriate C$_6$ coefficient of the considered pair.

Starting from a ultracold sample of Cs$_2$ molecules trapped in a QUEST, Feshbach-like collisions among molecules have been observed and interpreted as Cs$_4$ bound states \cite{chin2005}. In the experiment, the molecules are created from an atomic BEC in a crossed dipole trap by ramping the magnetic field through the g-wave Feshbach resonance at 19.84~G. A ramp of the magnetic field gradient levitates the molecules and separates them from the atoms, leaving a pure molecule cloud. Tuning the magnetic field further down, an avoided crossing at 13.6~G transfers the dimers from the initial state, with quantum numbers $f=4, m_f=4, \ell=4, m_{\ell}=2$, to another state ($f=6, m_f=6, \ell=4, m_{\ell}=0$). The new state has a different magnetic moment, so the transfer can be monitored by imaging the position of the sample that is due to a balance between magnetic force and gravity. In the new state two strong inelastic loss resonances are observed as a function of the magnetic field, at 12.72 and 13.15~G. These resonances are not reliable to the Cs$_2$ energy structure, that is well known near the dissociation limit until high partial waves \cite{chin2004a}, and can be explained as the occurrence of Cs$_4$ bound states. Measurements of the trap lifetime show a density-dependent effect that supports this interpretation.

An experiment on trapped heteronuclear molecules has investigated inelastic collisions as a function of the quantum statistics of the colliding particles \cite{zirbel2008}. KRb fermionic molecules were produced through rf-association near a Feshbach resonance between $^{40}$K$\left|9/2,-9/2\right\rangle$ and $^{87}$Rb$\left|1,1\right\rangle$ atoms, as in the experiment described in a previous section \cite{ospelkaus2006}. Once formed, the loss coefficient $\beta$ for inelastic collisions has been measured as a function of the heteronuclear scattering length \textit{a} for different collision partners and compared with theoretical behaviors \cite{dincao2006}; the investigated range was limited below 4500$a_0$, that is the region where the finite temperature does not cause dissociation. After removing Rb atoms, collisions with distinguishable atoms (K in the $\left|9/2,-7/2\right\rangle$ state) showed a \textit{a}$^{-1}$ dependence at high scattering length while at low \textit{a} $\beta$ stops changing. Collisions with bosons were studied in a mixture of R$b\left|1,1\right\rangle$ and  K$\left|9/2,-7/2\right\rangle$ atoms, by subtracting the contribution of K atoms. The loss coefficient shows in this case an increase at large \textit{a} due to the attractive atom-molecule interaction. Finally the molecule-fermionic atom collisions were studied by removing Rb atoms and driving K atoms to the $\left|9/2,-9/2\right\rangle$ state by rf. The results showed a suppression of the molecular decay at high \textit{a}. In this case, the trap lifetime $\tau$, that is related to the loss coefficient through $\beta=1/n\tau$, where n is the density of the atomic partner, reaches 100~ms.

Besides the investigation of cold collisions in trapped samples, it is now possible to manipulate the velocity of a molecular beam and thus to study processes and associated cross sections as a function of the collision energy. A first experiment of this kind has been reported in \cite{gilijamse2006}, where a molecular beam of OH in the $X^{2}\Pi_{3/2}, \nu=0, J=3/2, f$ state-selected state (where $f$ indicates the parity of the electronic wave function), after Stark deceleration, collided with a Xe beam in a crossed beam configuration. The velocity of the OH molecules was varied from 33 to 700~m.s$^{-1}$, with a low width of the velocity distribution, while the colliding Xe supersonic beam had a velocity of 300~m.s$^{-1}$ with a 10 \% spread that represented the main contribution to the energy resolution. Cross sections of inelastic scattering were measured as a function of the collision energy and compared with success to theoretical calculations for collisions changing the angular momentum and/or parity of the initial OH state. Another experiment probed collisions between trapped OH molecules and a supersonic beam of He or D$_2$ \cite{sawyer2008}. The OH molecules, after Stark deceleration, were trapped in a magnetic trap done with permanent magnets. The OH-He and OH-D$_2$ center-of-mass energy could be varied from 60 to 230~cm$^{-1}$ and from 145 to 510~cm$^{-1}$ respectively, by changing the nozzle temperature of the He or D$_2$ beam. Total collision cross sections have been measured by observing the OH population losses, showing for OH-He the signature of inelastic threshold like in the previous experiment \cite{gilijamse2006}, and evidence of resonant energy transfer in the case of OH-D$_2$ collisions.

\subsubsection{Modeling collisions between cold atoms and molecules}
\label{sssec:collisions}

The main limitation for such studies is again the lack of knowledge of the potential energy surfaces which drives the dynamical processes. Despite this difficulty, general trends can be identified, for instance by looking at the stability of the theoretical results with reasonable variations of the surfaces, as we will see below. Up to now, theoretical investigations covered several molecular systems, which are representative of the various cold or ultracold molecular systems available in experiments. Actually, theory cannot help too much to determine the magnitude of elastic cross sections, which crucially depend on the scattering length. Accurate spectroscopy of the related molecular systems is still lacking to determine this parameter. However, a wealth of theoretical works have focussed on low-energy collisions between atoms and molecules, as they imply the concept of Efimov states \cite{efimov1970}, which are a manifestation of a universal property of few-body systems. We will comment more about this in the next section, as they are the subject of a considerable literature.

Just like in the previous section, alkali compounds are attractive for modeling ultracold atom-molecule and molecule-molecule collisions, due to their simple structure, and as they are widely used in experiments. However, it is worthwhile to quote that computations for these effective three-electron systems are by far more demanding than those for the dimers. A systematic study of the potential energy surface of the lowest quartet state has been computed for all the homonuclear alkali trimers in their linear ($^4\Sigma^+$ state in $D_{\infty h}$ geometry) or equilateral configurations ($^4A'$ in $C_{3v}$ geometry), using standard {\it ab initio} packages like MOLPRO \cite{soldan2003}. The lowest quartet state is involved in a collision where all atoms are spin-polarized, which can be the case in realistic experiments, thus assuming that ultracold dimers are created in their lowest triplet state. The potential surfaces exhibit strong non-additive effects, i.e.  due to the strong polarizability of the alkali atoms, their equilibrium distance in the $C_{3v}$ symmetry is shorter than the one of the triplet dimer, and its energy minimum is lower than the sum of the depth of the pairwise potentials. Global potential energy surfaces have been subsequently derived on a grid of interatomic distances for Li$_3$ \cite{colavecchia2003,cvitas2005a}, Na$_3$ \cite{higgins2000,quemener2004}, and K$_3$ \cite{quemener2005}, in all geometries. New results have been recently obtained for heavier systems including heteronuclear alkali compounds like K$_i$Rb$_{3-i}$ ($i=1,3)$ in the context of molecules trapped onto helium nanodroplets \cite{hauser2008,nagl2008a}, or for the  Li$_2$A molecules (with A= Na, K, Rb, Cs) \cite{soldan2008}. As a general feature, all these surfaces have no potential barrier for atom exchange, which will be of central importance for modeling their dynamics. To be used in numerical codes for scattering, these surfaces have to be interpolated between the {\it ab initio} points, checking that no numerical oscillation arises, and that their dissociation limit is conveniently described. This requires specific techniques which are described for instance in ref.\cite{hutson2007}.

The simplest inelastic collision which can occur between a cold atom and a cold (diatomic) molecule is the change of the internal state of the molecule, like vibrational or relaxation defined as A+A$_2(v) \rightarrow $ A+A$_2  (v'<v)$. This process is of particular interest as in many experiments (like those of refs.\cite{staanum2006,zahzam2006,hudson2008,zirbel2008} involving alkali atoms) molecules are formed in high lying vibrational levels, in the presence of surrounding cold atoms. Despite its simplicity for alkali trimers, a full quantum dynamical treatment of the process, including inelastic and reactive channels, is required. Indeed, as no barrier is found in the relevant potential surfaces, an atom can be exchanged with the molecule (insertion mechanism) in the course of the collision involving indistinguishable alkali atoms. The choice of an appropriate set of internal coordinates is crucial to represent all possible arrangements simultaneously. Most models used hyperspherical coordinates $\rho, \theta, \phi$ which provide a global representation of the triatomic system: the hyperradius $\rho$ is related to the size of the triangle formed by the three particles, while $\theta$ and $\phi$ describe its shape. Various implementations have been proposed, which are reviewed for instance in ref.\cite{hutson2007}. In a series of studies, the dynamics of the A+A$_2$ collisions at ultracold energy (with A=Li, Na, K) has been investigated within the framework of a time-independent resolution of the Schr\"odinger equation written in hyperspherical coordinates, where A is spin-polarized in its ground state, and A$_2$ is in its lowest triplet state, possibly vibrationally excited. The first study of this kind has been performed on Na+Na$_2 (v=0,3)$ collisions \cite{soldan2002,quemener2004}, showing that vibrational relaxation of the molecule dominates the elastic scattering below 0.1~mK. This suggests that evaporative cooling of molecules by collisions with parent atoms is not favorable. Similar results have been obtained for Li+Li$_2 (v=0,3)$ \cite{cvitas2005a}, and for K+K$_2 (v=1)$ \cite{quemener2005}. The case of heteronuclear systems, involving mixed isotopes of the lithium atom has also been treated \cite{cvitas2005}, where the vibrational relaxation is found even stronger than for homonuclear case, compared to elastic processes. More recently a special care has been brought to the accurate determination of the long-range interactions implemented in the modeling of the Na+Na$_2$ and K+K$_2$ spin-changing collisions \cite{lysebo2008}. However, such calculations are still intended to provide a basic understanding of the processes, as they are restricted in several respects due to the limited capacity of computing facilities: the hyperfine structure and the presence of external fields are not included, and are restricted to low vibrational levels of the dimers, within the fully-spin-polarized configuration of the triatomic systems. The extension to other configurations is also of interest, as well as the treatment of heavier molecular systems or heteronuclar systems, providing that accurate potential surfaces could be calculated \cite{hauser2008,soldan2008}.

Even more desired by experimentalists are predictions concerning the interaction between a pair of ultracold alkali dimers. A full quantum mechanical study using a multidimensional potential surface is out of reach of the present computing resources. However some insight can already be obtained when a restricted region of the configuration space is examined. In ref.\cite{tscherbul2008}, the collision of pair of RbCs molecules has been investigated. The potential energy of the system has been calculated along a minimum energy path with an optimized geometry calculation. The authors showed that the reaction RbCs+RbCs$\rightarrow$ Rb$_2$+Cs$_2$ is barrierless, just like in the case of alkali atom-molecule collisions. In the spirit of the crossed molecular beam experiments of refs.\cite{elioff2003,strecker2008} on Ar-NO, and of ref.\cite{liu2007} on K-HBr, optimal conditions can be found to produce slow molecules from the reaction of two RbCs molecules.

Besides these studies on alkali compounds, numerous theoretical studies are devoted to the quantum description of inelastic or reactive scattering processes for other classes of systems. Such studies are actually progressing at the rhythm of the obtention of accurate potential energy surfaces, themselves motivated by the pressure of the increasing spread of experimental initiatives. Modeling inelastic collisions of molecules with Helium atoms at low energy looks promising in the perspective of the analysis of buffer-gas cooling experiments, as one of the collision partners (He) is structureless. Moreover, the Jacobi internal coordinate system - the molecular radius, the distance between the He atom and the center-of-mass of the molecule, and the angle between the molecular axis and the collision axis- is well adapted. The trapping efficiency of the cooled molecules critically depends on the Zeeman depolarization processes, i.e. of the relaxation of the molecules created in a well-defined Zeeman sublevel. Such processes are nicely reviewed in ref.\cite{bodo2006}, covering a variety of situations like open-shell or closed-shell molecular systems, and relaxation processes either from low or high-lying rovibrational levels. Among them, we can quote the treatment of the collisions of He with N$_2$ \cite{stoecklin2002}, with CaH \cite{balakrishnan2003}, with NH \cite{cybulski2005}, with CO \cite{yang2005}, with OH \cite{gonzalez-sanchez2006}, or with Li$_2$ \cite{bodo2006a}, all species which have been cooled down in their ground state. All results demonstrate a strong influence of the initial internal state of the molecule on the relaxation cross section.

\begin{figure}[htb]
\begin{center}
\includegraphics[width=10cm,height=8cm]{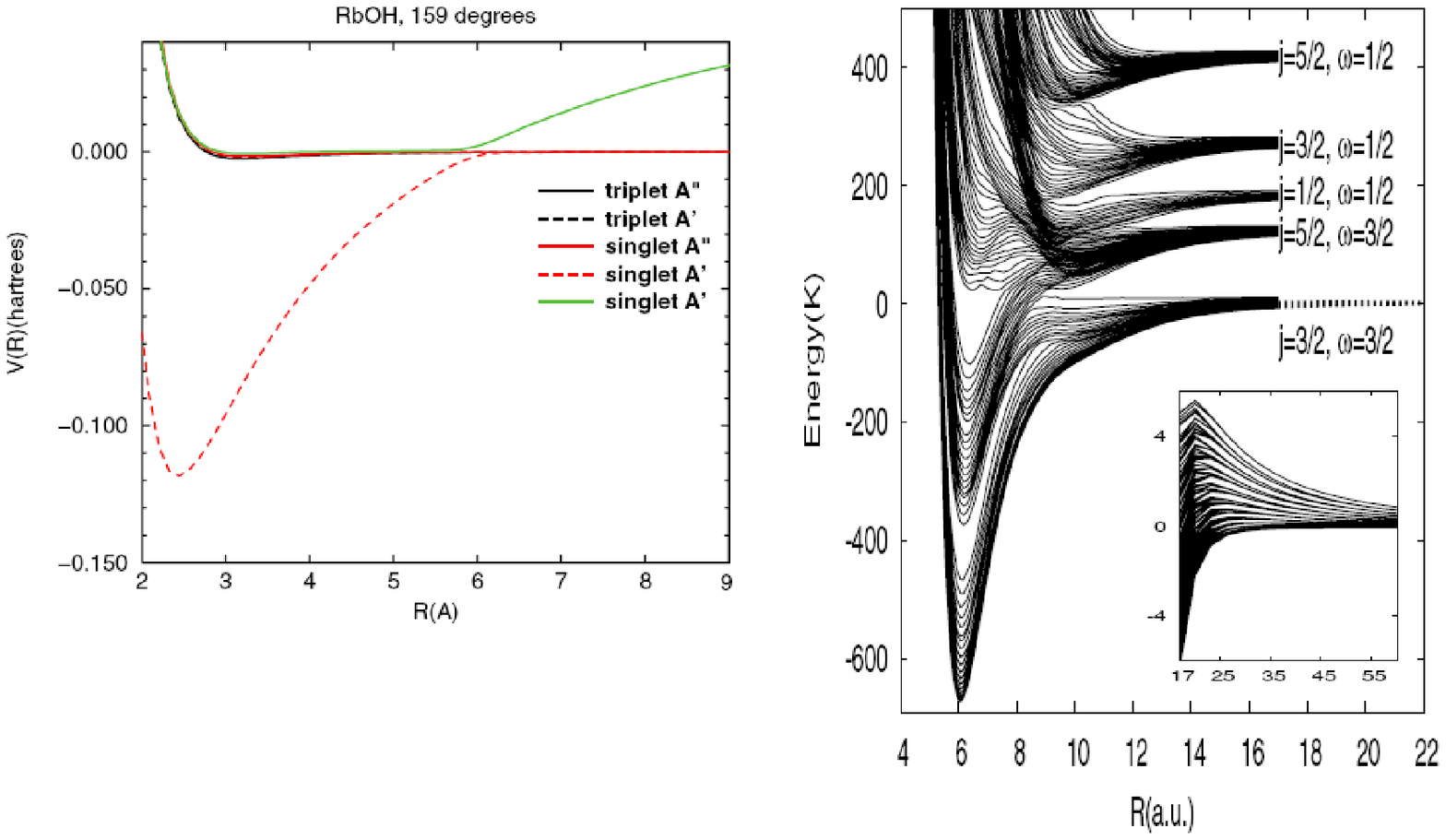}
\end{center}
\caption{Potential energy surfaces of the OH-Rb system. Left: adiabatic potential curves showing crossing for the avoided crossing between the ionic and covalent channels for a slightly nonlinear geometry. Right: adiabatic curves correlating with the lower rotational states for the collision in maximally stretched states of both partners, including hyperfine interactions. Reprinted with permission  Lara {\it et al} \cite{lara2007}.}
\label{fig:rboh}
\end{figure}

In a couple of other studies, the possibility to use ultracold alkali atoms to sympathetically cool down molecules obtained from devices like Stark decelerator is investigated. Compared to the case above, both reactants are open-shell systems, which represents a challenge for {\it ab initio} calculations as several coupled potential surfaces have to be determined. The interaction between the polar NH($X^3\Sigma^-)$ molecule and Rb or Cs atoms has been studied in \cite{tacconi2007,tacconi2007a}, and between OH($X^2\Pi$) and Rb atoms in \cite{lara2006,lara2007}, both including spin-orbit effects. One particular feature of such systems is the presence of an ionic channel, i.e. Rb$^+$+NH$^-$ or Rb$^+$+OH$^-$, whose energy at infinity is not far above the one of the neutral system. As a consequence, the interaction energy drops very fast when the mutual distance decreases, and generates a conical intersection between the potential energy surface of this ionic channel and the one of the neutral system at short distances (Fig. \ref{fig:rboh}a). Hyperfine interaction has been included in the latter case, which tremendously increases the number of channels in the scattering process (Fig. \ref{fig:rboh}b). In all these studies, inelastic collisions which change the internal state of the molecules dominate elastic ones, which could make sympathetic cooling problematic for trapped molecules.

Following the path of increasing complexity, several explorative studies have been devoted to cold collisions of identical molecules. The H$_2$+H$_2$ system is obviously a prototypical one, being the simplest tetramer system. These investigations take advantage of the improvement of potential energy surfaces, as well as of the continuous increase of numerical facilities. Analysis of rotational energy transfer mechanisms has been carried out over a wide range of energies including the ultracold range, examining also the dependence of the results with the choice of the potential energy surface \cite{lee2006,quemener2008,quemener2009}. Other studies concern collisions between O$_2$ molecules in various internal states \cite{avdeenkov2001,tilford2004}. It is important to note that the perspective of such works is the modeling of collisions taking place inside molecular traps, which are, by construction, created by external electric and/or magnetic fields. Again, as discussed above, these fields change the properties of the molecules, and therefore influence their dynamics especially at low energy. This opens the way to the control of elementary chemical reactions at low energy using external fields, which is the topic of two recent detailed reviews \cite{krems2005,krems2008}.

\section{Concluding remarks: accessing new phenomena with cold molecules, a challenge for physicists and chemists.}

In this review, we have presented the huge advances in the field of ultracold molecules, addressing the main experimental and theoretical issues concerned by their formation process. The dream of controlling both the internal and the external degrees of freedom of a molecule seems now to be at reach. Photoassociation, either followed by spontaneous emission \cite{deiglmayr2008} or combined with incoherent optical techniques \cite{sage2005,viteau2008}, can produce ultracold dimers in the ground rovibrational state with a temperature in the $\mu$k range. Ultracold molecules in the ground rovibrational level at even lower temperatures (in the nK range) can now be produced by coherent optical transfer of magnetoassociated molecules \cite{lang2008a,danzl2008a,ni2008}. Quantum degeneracy of a molecular gas, already demonstrated for Feshbach molecules \cite{jochim2003a,zwierlein2003,greiner2003}, will presumably be achieved soon also for molecules in the ground rovibrational state. These association techniques are currently limited to alkali species, but different kinds of molecules can be cooled using methods like Stark or Zeeman deceleration and kinematic cooling. Such techniques create molecular samples in the mK temperature range, but there are proposals to further cool and compress these molecules in the phase space \cite{stuhl2008}.

The broad area of physical and chemical applications of ultracold molecules, that covers ultra-high resolution spectroscopy with tests of fundamental theories, few-body physics, quantum computation, molecular optics and controlled chemical processes, is now beyond the proposal stage and begins to show the first important results. Such an exhaustive presentation of these topics would require a review twice long, so that we summarize in this last section their most promising or spectacular developments.

\begin{itemize}
\item {\it Novel ultra-high resolution molecular spectroscopy}

The resolution of conventional molecular spectroscopy benefits of the utilization of cold molecular samples, producing a strong reduction of first and second order Doppler effect and a longer interrogation time in the measuring device. But the main advantages of cold molecules rely in unconventional spectroscopy, as the access to pure long-range states and to triplet states is hard to achieve with thermal ensembles. As described in section \ref{sssec:pa}, PA allows to obtain detailed information on the long-range part of molecular potential curves and to the properties of the constituting atoms. A huge amount of accurate atomic and molecular data have been reported in literature in the last two decades \cite{weiner1999,stwalley1999,jones2006}. An example is the scattering length $a$, already quoted in section \ref{sssec:magnetoass}, that deserves very precise determination. As it is well known, this parameter resumes the elastic scattering properties of ground state atoms at low collision energies, and plays a crucial role in the stability of Bose-Einstein condensates. The scattering length is related to the total accumulated phase of the interaction potential between two atoms, and is extremely sensitive to any detail of this potential at a level which cannot be achieved by pure theoretical approaches. Instead joint experimental and theoretical studies have been amazingly successful to determine this parameter, and to predict the strongly coupled energy level structure of long-range molecules induced by hyperfine interactions. PA can provide good estimates of $a$ by locating the nodal positions of the wave function or by measuring the near-threshold bound states. The novel {\it Feshbach spectroscopy} is highly accurate in measuring near-threshold bound states in all systems that exhibit Feshbach resonances. By elaborate time-dependent control of the magnetic field it is possible to populate any of the near-threshold bound molecular states \cite{mark2007}, including those having high rotational angular momentum \cite{knoop2008}. The states that cannot be directly populated from the open channel can be reached through state transfer, with either a slow (adiabatic) or fast (diabatic) field ramp through a level crossing. Spectroscopic signature of a level crossing can be given by a measurement of the molecular magnetic moment or by dissociation loss after applying microwave radiation \cite{mark2007}. Also transfer across avoided crossings using radiofrequency allowed cruising through the Feshbach resonance manifold of $^{87}$Rb$_2$ \cite{lang2008}. Several authors proposed broad reviews of the physics of Feshbach resonances, and of their applications \cite{timmermans1999,kohler2006,chin2009}.

\item{\it Tests of fundamental theories}

The ultra-high resolution which is made available in cold molecule experiments allows to design protocols aiming to perform tests of fundamental theories like the cosmological variation of the fine structure constant $\alpha$ \cite{veldhoven2004} or the $m_e/m_p$ ratio \cite{karr2005,schiller2005,chin2006,demille2008,kajita2008,zelevinsky2008}. One example is the measurement of the $\Lambda$-doublet transitions of OH ground-state molecules in the microwave range \cite{hudson2006a}. A beam of OH molecules was Stark-decelerated before entering a microwave cavity where the molecules were probed by Rabi and Ramsey spectroscopy, giving a large improvement over the previous frequency measurements thanks to the longer interrogation time. The sum and difference of $\Delta$F=0 transition frequencies can be related to the fine structure constant $\alpha$, and a comparison of laboratory and astrophysical measurements can give constraints on the time variation of $\alpha$, that is predicted by unified field theories \cite{uzan2003}.

The measurement of the electric dipole moment (EDM) of the electron is another example where cold molecules can be useful. A non-zero EDM would imply a T-violation, i.e. an asymmetry with respect to time reversal, and a limit value would test theories beyond the standard model \cite{hunter1991}. A heavy polar molecule is a good candidate for a sensitive measurement, as the effective electric field felt by the unpaired electron can be expressed by the product $Q \times P$, where $P$ is the molecular polarizability induced by an external field ($\approx$1 for a polar molecule) and $Q$ is a factor proportional to $Z^3$, with $Z$ the mass number \cite{demille2000}. Experiments aiming at determining the electron EDM are underway on PbO \cite{egorov2001,kawall2004,kawall2005,bickman2008} and YbF \cite{hudson2002,tarbutt2004,tarbutt2009}.

\item {\it The quantum degenerate regime for molecules}

One of the most important result in the field has been the achievement of a BEC of Feshbach molecules out of a degenerate atomic Fermi gas (see section \ref{sssec:magnetoass}). Besides the interest in obtaining a quantum degenerate regime for molecules in the ground rovibrational state, i.e. without internal energy, both theory and experiments have deeply investigated the region very close to a Feshbach resonance. The region where the scattering length $a<0$ does not support bound molecular states according to two-body physics. On the contrary, many-body physics, in the limit of weak interactions, allows for the formation of fermionic pairs identifiable with Cooper pairs. Following the Bardeen-Cooper-Schrieffer (BCS) theory \cite{bardeen1957}, that explains the occurrence of superconductivity and superfluidity, Cooper pairs are delocalized pairs of electrons weekly bound at very long range. Bose-Einstein condensation of Cooper pairs was observed in both K and Li gases by a rapid sweep of the magnetic field from the BCS to the BEC side of the resonance \cite{regal2004a,zwierlein2004} and the pairing energy gap, as forecast by the theory, has been detected \cite{chin2004,greiner2005,partridge2005}.
The BEC and BCS regions are connected by a crossover region where the gas is strongly interacting and theory becomes a challenging problem \cite{bloch2008}.

Starting from an atomic Bose-Einstein condensate, the coherent conversion into a molecular BEC is an advance towards a sort of "{\it superchemistry}" that is a stimulated emission of molecules in a chemical reaction \cite{heinzen2000}. Nonlinearity in the atom-molecule coupling through stimulated Raman transitions may give rise to giant collective oscillations with time-dependent reaction rates, that depend on the reaction vessel. The atom-molecule system can present quantum effects like soliton formation or squeezed state generation.

\item {\it Universality of few-body physics at ultracold temperatures}

In the ultracold domain, the scattering length is the characteristic parameter of the mutual interaction between atoms, controlling their elastic collisions. If the scattering length reaches large values (for instance by tuning it with external magnetic fields), it can exceed the range of molecular interaction (halo system), so that the long-range atom-atom interaction becomes independent of any detail of the short-range interaction. For instance, the binding energy $E_b$ of the uppermost bound state of an atom pair characterized by a positive scattering length $a$ is given by $E_b=\hbar^2/(\mu a^2)$, where $\mu$ is the reduced mass of the system. This property suggests that two-body properties at ultracold energies adopt a universal character.

Such a universality has been demonstrated for three-body systems by Efimov \cite{efimov1970}. In the limit of infinite scattering length, he demonstrated that a series of three-body bound states converges towards the dissociation threshold, with ratio of binding energies of successive states equal to a universal constant. Efimov states  actually occur in three identical boson systems in presence of a two-body interaction with large scattering length \cite{efimov1970}. The so-called Efimov physics has been reviewed in depth by Braaten and coworkers \cite{braaten2006,braaten2007}.

The availability of ultracold samples of atomic and molecular particles, together with the possibility to tune their mutual interactions through Feshbach resonances opened new perspectives for the investigation of Efimov physics, illustrated by two recent experiments bringing first evidences of Efimov physics. In ref.\cite{kraemer2006}, a resonance in the three-body recombination loss rate measured in a cesium trap has been observed in the region of large negative (two-body) scattering length near a Feshbach resonance, yielding an indirect manifestation of an Efimov state. An atom-dimer resonance has also been observed in an optically-trapped mixture of Cs atoms and Cs$_2$ Feshbach molecules \cite{knoop2008a}, suggesting the presence of a trimer state close to the atom-dimer dissociation threshold. This exploration has been extended to the collisions between Feshbach Cs$_2$ molecules in the quantum halo regime, measuring their decay rate down to lower vibrational states \cite{ferlaino2008}. The authors observed a pronounced minimum in the loss rate when varying the scattering length, which has still to be interpreted in relation with universal properties of four-body physics.

\item {\it Dipolar gases}

A novel aspect of ultracold atomic and molecular gases emerged with the realization of samples of ultracold particles with a permanent electric dipole moment such as chromium atoms, or heteronuclear alkali diatoms, or with an electric dipole moment induced by an external electric field, like cold Rydberg atoms. Indeed, such particles interact through long-range dipole-dipole potentials varying as $R^{-3}$ (with $R$ their mutual distance) which dominates the usual van der Waals interaction (varying as $R^{-6}$). Such cold gases will evolve in a entirely new regime of strong interactions with pronounced anisotropy, compared to previous studies. A considerable literature is now available treating many aspects of such dipolar gases \cite{baranov2007,lewenstein2007,pupillo2008}, whose interaction between particles could be controlled either by tuning external fields, or by using various geometries of the trapping configuration, like quasi-2D traps for instance. In particular, dipolar gases represent systems of choice for many-body physics, providing new possibilities for modeling condensed matter phases.

\item {\it Quantum information}

The development of a quantum computer, that should perform some tasks with exponentially higher efficiency than a classical computer, is actually being pursued. Neutral polar molecules are suitable as information carriers because they have well-defined internal states, they weakly interact with the environment, and they can be entangled through dipole-dipole interaction. A first proposal to use cold polar molecules for this purpose was given in \cite{demille2002}. The polar molecules should be trapped in a 1D array and oriented along or against an electric field; it is remarkable that the loading of a single molecule in each lattice site has already been experimentally demonstrated \cite{volz2006}. An electric field gradient allows for spectroscopic addressing of a single molecule, through microwave field pulses that couple rotational states, implementing the quantum information processing.  Using a 1D array of KCs dimers (or similar) in the ground rovibrational state, about 10$^5$ CNOT gates should be feasible within the coherence time of a few seconds. Several scenarios have been proposed using ultracold polar molecules to achieve entanglement tests \cite{milman2007}, quantum phase gates \cite{charron2007}, or robust quantum computation schemes \cite{yelin2006,kuznetsova2008}. In order to combine the coherence properties of an ensemble of cold polar molecules together with a scalable architecture, a quantum computer using molecules and a mesoscopic solid-state resonator has been proposed \cite{andre2006}. Trapping of cold polar molecules like CaBr above an electrostatic Z-trap, integrated with a microwave strip-line resonator, would induce also a cooling process of the molecules through cavity-enhanced spontaneous emission of excited rotational states. The process would be similar to sideband cooling used for trapped ions \cite{wineland1979}. Finally the utilization of cold polar molecules as quantum memories of solid-state circuits have been proposed \cite{rabl2006,rabl2007}.

\end{itemize}

\section*{Acknowledgments}
We are indebted to Daniel Comparat, Goran Pichler, and Marin Pichler, for their critical reading of the manuscript. This work is supported in part by the network "Quantum Dipolar Molecules (QuDipMol)" of the EUROQUAM framework of the European Science Foundation (ESF).

\newpage

\section*{References}

\bibliography{../DATABASE/bibliocold}
\bibliographystyle{iopart-num}

\providecommand{\newblock}{}

\end{document}